\documentclass[useAMS,referee,usenatbib]{biom}
\def\bSig\mathbf{\Sigma}

 \renewcommand{\theequation}{\thesection.\arabic{equation}}

\usepackage{amsmath, amssymb, amsfonts, amsbsy,  epsfig, graphicx,rotating, subfigure}
\usepackage[usenames]{color}
\usepackage{enumerate}
\usepackage{natbib}

\RequirePackage[colorlinks,linkcolor=red,citecolor=blue,urlcolor=magenta]{hyperref}
 \usepackage{ulem}
\usepackage{multirow,xspace,setspace,booktabs,subfigure}
 \usepackage{fancyhdr}
\usepackage{tabularx}
\usepackage{tabulary}
\usepackage{caption}
\usepackage{stmaryrd}

\numberwithin{equation}{section}
\def\be{\begin{equation}}
\def\ee{\end{equation}}
\def\bea{\begin{eqnarray}}
\def\eea{\end{eqnarray}}
\def\bd{\begin{displaymath}}
\def\ed{\end{displaymath}}
\def\bda{\begin{eqnarray*}}
\def\eda{\end{eqnarray*}}
\def\bsm{\begin{small}}
\def\esm{\end{small}}

\def\t0{\theta_0}

\def\nn{\nonumber}

\def\ha1{\hat \beta_1}

\def\bnt{\begin{enumerate}}
\def\ent{\end{enumerate}}
\def\T{{ \mathrm{\scriptscriptstyle T} }}

\def\bsc{\begin{scriptsize}}
\def\esc{\end{scriptsize}}

\newtheorem{ft}{Fact}[section]
\theoremstyle{definition}

\newcommand{\wh}{\widehat}

\makeatletter
\newcommand{\figcaption}{\def\@captype{figure}\caption}
\newcommand{\tabcaption}{\def\@captype{table}\caption}
\makeatother

 \newcommand{\var}{\mbox{$\mathbb{V}$ar}}

\def\ga{\gamma}
\def\de{\delta}

\def\la{\lambda}

\newcommand{\sn}{\sum_{i=1}^{n}}
\newcommand{\sump}{\sum_{j=1}^m}

\renewcommand{\P}{\mathbb{P}}

\newcommand{\br}{\mathbb{R}}

\newcommand{\bA}{{\mathbf A}}

\newcommand{\bF}{{\mathbf F}}

\newcommand{\bI}{{\mathbf I}}

\newcommand{\bQ}{{\mathbf Q}}

\newcommand{\bR}{{\mathbf R}}

\newcommand{\bU}{{\mathbf U}}

\newcommand{\bX}{{\mathbf X}}
\newcommand{\bY}{{\mathbf Y}}
\newcommand{\bT}{{\mathbf T}}
\newcommand{\bZ}{{\mathbf Z}}

\newcommand{\bSigma}{\boldsymbol{\Sigma}}

\newcommand{\bmu} {\boldsymbol{\mu}}

\newcommand{\bD}{{\mathbf D}}

\newcommand{\beq}{\begin{eqnarray*}}
\newcommand{\eeq}{\end{eqnarray*}}
\newcommand{\beqn}{\begin{eqnarray}}
\newcommand{\eeqn}{\end{eqnarray}}

\newcommand{\sm}{\sum_{j=1}^m}

\def\T{{ \mathrm{\scriptscriptstyle T} }}

\title[Testing Large Covariance Matrices] {Comparing Large Covariance Matrices under Weak Conditions on the Dependence Structure and its Application to Gene Clustering}
\author{Jinyuan Chang$^{1,2,*}$\email{jinyuan.chang@unimelb.edu.au},
Wen Zhou$^{3,**}$\email{riczw@stat.colostate.edu}, Wen-Xin Zhou$^{4,***}$\email{wenxinz@princeton.edu}, and Lan Wang$^{5,****}$\email{wangx346@umn.edu}  \\
\small $^{1}$School of Statistics, Southwestern University of Finance and Economics, Chengdu, Sichuan 611130, China\\
\small $^{2}$School of Mathematics and Statistics, The University of Melbourne, Parkville, VIC 3010, Australia \\
\small $^{3}$Department of Statistics, Colorado State University, Fort Collins, CO 80523, U.S.A.
\\
\small $^{4}$Department of Operations Research and Financial Engineering, Princeton University, Princeton, NJ 08544, U.S.A.\\
\small $^{5}$School of Statistics, University of Minnesota, Minneapolis, MN 55455, U.S.A.}

\begin{document}

%  This will produce the submission and review information that appears
%  right after the reference section.  Of course, it will be unknown when
%  you submit your paper, so you can either leave this out or put in
%  sample dates (these will have no effect on the fate of your paper in the
%  review process!)

\date{{\it Received September} 2015.  }

%  These options will count the number of pages and provide volume
%  and date information in the upper left hand corner of the top of the
%  first page as in published papers.  The \pagerange command will only
%  work if you place the command \label{firstpage} near the beginning
%  of the document and \label{lastpage} at the end of the document, as we
%  have done in this template.

%  Again, putting a volume number and date is for your own amusement and
%  has no bearing on what actually happens to your paper!

\pagerange{\pageref{firstpage}--\pageref{lastpage}}
\volume{}
\pubyear{2015}
\artmonth{September}

%  The \doi command is where the DOI for your paper would be placed should it
%  be published.  Again, if you make one up and stick it here, it means
%  nothing!

\doi{10.1111/j.1541-0420.2005.00454.x}

%  This label and the label ``lastpage'' are used by the \pagerange
%  command above to give the page range for the article.  You may have
%  to process the document twice to get this to match up with what you
%  expect.  When using the referee option, this will not count the pages
%  with tables and figures.

\label{firstpage}

%  put the summary for your paper here

\begin{abstract}
{Comparing large covariance matrices has important applications in modern genomics, where scientists are often interested in understanding whether relationships (e.g., dependencies or co-regulations) among a large number of genes vary between different biological states. We propose a computationally fast procedure for testing the equality of two large covariance matrices when the dimensions of the covariance matrices are much larger than the sample sizes. A distinguishing feature of the new procedure is that it imposes no structural assumptions on the unknown covariance matrices. Hence the test is robust with respect to various complex dependence structures that frequently arise in genomics. We prove that the proposed procedure is asymptotically valid under weak moment conditions. As an interesting application, we derive a new gene clustering algorithm which shares the same nice property of avoiding restrictive structural assumptions for high-dimensional genomics data. Using an
asthma gene expression dataset, we illustrate how the new test helps compare the covariance matrices of the genes across different gene sets/pathways between the disease group and the control group, and how the gene clustering algorithm provides new insights on the way gene clustering patterns differ between the two groups. The proposed methods have been implemented in an \texttt{R}-package \texttt{HDtest} and is available on CRAN.}
\end{abstract}


%  As usual, the \maketitle command creates the title and author/affiliations
%  display

\maketitle

%  If you are using the referee option, a new page, numbered page 1, will
%  start after the summary and keywords.  The page numbers thus count the
%  number of pages of your manuscript in the preferred submission style.
%  Remember, ``Normally, regular papers exceeding 25 pages and Reader Reaction
%  papers exceeding 12 pages in (the preferred style) will be returned to
%  the authors without review. The page limit includes acknowledgements,
%  references, and appendices, but not tables and figures. The page count does
%  not include the title page and abstract. A maximum of six (6) tables or
%  figures combined is often required.''

%  You may now place the substance of your manuscript here.  Please use
%  the \section, \subsection, etc commands as described in the user guide.
%  Please use \label and \ref commands to cross-reference sections, equations,
%  tables, figures, etc.
%
%  Please DO NOT attempt to reformat the style of equation numbering!
%  For that matter, please do not attempt to redefine anything!

%=================================================================================================%

\section{Introduction}
\label{intro}
The problem of comparing two large population covariance matrices has important applications in modern genomics,
where growing attentions have been devoted to understanding how the relationship (e.g. dependencies or co-regulations) among genes
vary between different biological states. Our interest in this problem is motivated by a microarray study on human asthma \citep{V2014}. This study consists of 88 asthma patients and 20 controls.  It is known that genes tend to work collectively
in groups to achieve certain biological tasks. Our analysis focuses on such groups of genes (gene sets) defined with the gene ontology (GO) framework, which are referred to as GO terms. Identifying GO terms with altered dependence structures between disease and control groups  provides critical information on differential gene pathways associated with asthma. Many of the GO terms contain a large number of (in the asthma data, as many as 8,070) genes. The large dimension of microarray data and the complex dependence structure among genes make the problem of comparing two population matrices extremely challenging.

In conventional multivariate analysis where the dimension $p$ is fixed, testing the equality of two unknown covariance matrices $\bSigma_1$ and $\bSigma_2$ based on the samples with sample sizes $n$ and $m$ has been extensively studied, see for example \cite{Anderson_2003} and {the} references therein. In the high-dimensional setting where $p>\max(n,m)$, recently several authors have developed new tests other than the traditional likelihood ratio test. Considering multivariate normal data, \cite{Schott_2007} and \cite{SrivastavaYanagihara_2010} constructed tests using different distances based on traces of the covariance matrices; \cite{LiChen_2012} proposed a $U$-statistic based test for a more general multivariate model. These tests are effective for dense alternatives, but often suffer from low power when $\bSigma_1-\bSigma_2$ is sparse. We are more interested in this latter situation, as in genomics the difference in the dependence structures between populations typically involves only a small number of genes.
%when we compare the dependence structures of genes between disease and control groups,

For sparse alternatives, \cite{CaiLiuXia_2013} investigated an $L_\infty$-type test. They proved that
the distribution of the test statistic converges to a type I extreme value distribution under the null hypothesis and
the test enjoys certain optimality property. Motivated by this work, we propose in this paper a perturbed variation of
the $L_\infty$-type test statistic. We verify that the conditional distribution of the  perturbed  $L_\infty$-statistic
provides a high-quality approximation to the distribution of the original  $L_\infty$-type test,
which has important implications in achieving accurate performance in finite
sample size. In contrast, the convergence rate to the extreme-value distribution of type I is of order $O\{ \log(\log n)/\log n\}$ \citep{LiuLinShao_2008}.

The asymptotic validity of our proposed new procedure does not require any structural assumptions on the unknown covariances. It is valid under weak moment conditions.
On the other hand, the aforementioned work all require certain parametric distributional assumptions or structural assumptions on the population covariances in order to derive
an asymptotically pivotal distribution.
Assumptions of this kind are not only difficult to be verified but also often violated in real data.
It is known that expression levels of the genes regulated by the same pathway \citep{WM2012} or associated with the same functionality \citep{K2014} are often highly correlated. Also, in the microarray and sequencing experiments, most genes are expressed at very low levels while few are expressed at high levels. This implies that the distribution of gene expressions is most likely heavy-tailed regardless of the normalization and transformations \citep{WPL2015}.

%Specifically, we consider testing $H_0: \bSigma_1 = \bSigma_2$ in high dimensions against sparse alternatives.

For testing $H_0: \bSigma_1 = \bSigma_2$ in high dimensions, the new procedure is computationally fast and adaptive to the unknown dependence structures. Section \ref{method.sec} introduces the new testing procedure and investigates its theoretical properties. In Section \ref{simulation.sec}, we compare its finite sample performance with several competitive procedures. A gene clustering algorithm is derived in Section \ref{sec51}, which aims to group hundreds or thousands of genes based on the expression patterns \citep{SES2002} without imposing restrictive structural assumptions. We apply the proposed procedures to the human asthma dataset in Section \ref{real}. Section \ref{discuss} discusses our results and other related work. Proofs of the theoretical results and additional numerical results are provided in the Supplementary Material. The proposed methods have been implemented in the \texttt{R} package \texttt{HDtest} and is currently available on CRAN (\url{http://cran.r-project.org}).

\section{The new testing procedure}
\label{method.sec}

\subsection{The $L_\infty$-statistic}
%\label{sec21}

Let $\bX = (X_1, \ldots, X_p)^{\T}$ and $\bY = ( Y_1, \ldots, Y_p)^{\T}$ be two $p$-dimensional random vectors with means $\bmu_1=(\mu_{11},\ldots,\mu_{1p})^{\T}$ and $\bmu_2=(\mu_{21},\ldots,\mu_{2p})^{\T}$, and covariance matrices $\bSigma_1{=(\sigma_{1,k\ell})_{1\leq k,\ell \leq p}}$ and $\bSigma_2{=(\sigma_{2,k\ell})_{1\leq k,\ell \leq p}}$, respectively. We are interested in testing
\be
     H_0: \bSigma_1 = \bSigma_2 ~~~~\text{versus}~~~~ H_1: \bSigma_1 \neq \bSigma_2 \label{covariance.test}
\ee
based on independent random samples $\mathcal{X}_n=\{\bX_1,\ldots, \bX_n\}$ and $\mathcal{Y}_m=\{\bY_1,\ldots, \bY_m\}$ drawn from the distributions of $\bX$ and $\bY$, respectively. For each $i$ and $j$, we write $\bX_i=(X_{i1},\ldots, X_{ip})^{\T}$ and $\bY_j = (Y_{j1},\ldots, Y_{jp})^{\T}$. Let $\wh{\bSigma}_1=(\hat{\sigma}_{1,k\ell})_{1\leq k,\ell \leq p} =  n^{-1}\sn (\bX_i -\bar{\bX} ) (\bX_i-\bar{\bX})^{\T}$ and $\wh{\bSigma}_2=(\hat{\sigma}_{2,k\ell})_{1\leq k,\ell \leq p} =  m^{-1}\sum_{j=1}^m (\bY_j -\bar{\bY}) (\bY_j - \bar{\bY})^{\T}$ be the sample analogues of $\bSigma_1$ and $\bSigma_2$, where $\bar{\bX}=(\bar{X}_1,\ldots,\bar{X}_p)^{\T}=n^{-1}\sn \bX_i$ and $\bar{\bY}=(\bar{Y}_1,\ldots, \bar{Y}_p)^{\T}=m^{-1}\sm \bY_j$.

For each $(k,\ell)$, a straightforward extension of the two-sample $t$-statistic
for the marginal {hypothesis} $H_{0,k\ell}:\sigma_{1,k\ell}=\sigma_{2,k\ell}$ versus $H_{1,k\ell}:\sigma_{1,k\ell}\neq\sigma_{2,k\ell}$
is given by
\be
    \hat{t}_{k\ell}= \frac{ \hat{\sigma}_{1,k\ell}-\hat{\sigma}_{2,k\ell} }{ ( n^{-1} \hat{s}_{1,k\ell} + m^{-1} \hat{s}_{2,k\ell} )^{1/2}},  \label{eq2.3}
\ee
where $ \hat s_{1,k\ell} = n^{-1} \sn  \{  (X_{ik}-\bar{X}_k)(X_{i\ell}-\bar{X}_\ell ) - \hat{\sigma}_{1,k\ell}  \}^2$ and $\hat s_{2,k\ell} = m^{-1} \sump \{ (Y_{jk}-\bar{Y}_k ) (Y_{j\ell} -\bar{Y}_\ell) - \hat \sigma_{2,k\ell}  \}^2 $ are estimators of $s_{1,k\ell}=\var\{(X_{k}-\mu_{1k})(X_{\ell} -\mu_{1\ell})\}$ and $s_{2,k\ell}=\var\{(Y_{k}-\mu_{2k})(Y_{\ell} -\mu_{2 \ell})\}$, respectively.

Since the null hypothesis in (\ref{covariance.test}) is equivalent to {$H_0: \max_{1\leq k\leq \ell \leq p} | \sigma_{1,k\ell}- \sigma_{2,k\ell} | =0$},
a natural test statistic that is powerful against sparse alternatives in (\ref{covariance.test}) is the $L_\infty$-statistic
\begin{align}
    \wh{T}_{\max} = \max_{1\leq k \leq \ell \leq p}  | \hat t_{k\ell} |.  \label{Tmax}
\end{align}

\subsection{A new testing procedure}
\label{sec22a}
One way to base a testing procedure on the $L_\infty$-statistic is to reject the null hypothesis (\ref{covariance.test}) when $\wh{T}_{\max}^2-4\log p+\log(\log p)>q_{\alpha}$, where $q_{\alpha}=-\log(8\pi)-2\log\log (1-\alpha)^{-1}$ corresponds to the $(1-\alpha)$-quantile of the type I extreme value distribution. \cite{CaiLiuXia_2013} proved that this leads to a test that maintains level $\alpha$ asymptotically and enjoys certain optimality.

In this section, we propose a new test that rejects  (\ref{covariance.test}) when $\wh{T}_{\max}>{c}_{\alpha}$,
where ${c}_{\alpha}$ is obtained using a fast-computing data perturbation procedure.
The new procedure resolves two issues at once. First, it achieves better finite sample performance by avoiding the
slow convergence of $\wh{T}_{\max}^2-4\log p+\log(\log p)$ to the  type I extreme value distribution. Second and more importantly,
our procedure relaxes the conditions on the covariance matrices required in  \cite{CaiLiuXia_2013} (particularly, their Conditions (C1) and (C3)). %, which are hard to verify for real data analysis.
Note that their Condition (C1) essentially requires that
the number of variables that have non-degenerate correlations with others should grow no faster
than the rate of $p$. Although this condition is reasonable in some applications, it is
hard to be justified for data from the microarray or transcriptome experiments, where the genes can be divided into gene sets with varying sizes according to functionalities,
and usually genes from the same set have relatively high (sometimes very high) intergene correlations compared to those from different sets. %while genes from different pathways have relatively weak correlation.
This corresponds to an approximate block structure. Many sets can contain several thousand genes,
a polynomial order of $p$. This kind of block structure with growing block size may violate
Condition (C1) in \cite{CaiLiuXia_2013}. The crux of the derivation of the asymptotic type I extreme value distribution
in \citep{CaiLiuXia_2013}
is that the $\hat t_{k\ell}$'s are weakly dependent under $H_0$ under certain regularity conditions. In contrast,
the new procedure we present below automatically takes into account correlations among the $\hat{t}_{k\ell}$'s.

%Before describing the new procedure, we first provide some intuition to the rationale. Under %$H_{0,k\ell}$, it follows from the central limit theorem that under proper regularity conditions,
%$\hat{t}_{k\ell}=(n^{-1}s_{1,k\ell}+m^{-1}s_{2,k\ell})^{1/2}(\hat{\sigma}_{1,k\ell}-\hat{\sigma}_{2,k\ell})+ {O_{{\rm p}}}(m^{-1/2}+n^{-1/2}) .$ This suggests that the dependency between $\hat{t}_{k_1\ell_1}$ and $\hat{t}_{k_2\ell_2}$ under $H_0$ in (\ref{covariance.test}) with distinct indices can be approximated by that between their leading terms under the null. Notice that the leading terms of $\hat{t}_{k_1\ell_1}$ and $\hat{t}_{k_2\ell_2}$ behave like normal random variables, hence the dependency between $\hat{t}_{k_1\ell_1}$ and $\hat{t}_{k_2\ell_2}$ can be approximately evaluated by the covariance/correlation between their leading terms.

Specifically, we propose the following procedure to {compute} $c_\alpha$ with the dependence among $\hat{t}_{k\ell}$'s incorporated.

(I). Independent of $\mathcal{X}_n$ and $\mathcal{Y}_m$, we generate a sequence of independent $N(0,1)$ random variables $ g_{1},\ldots, g_{N}$, where $N=n+m$ is the total sample size.

(II). Using the $g_i$'s as multipliers, we calculate the perturbed version of the test statistic
\be
    \wh{T}^\dagger_{\max} = \max_{1\leq k\leq \ell \leq p}  |\hat{t}_{k\ell}^{\, \dagger}| , \label{eq2.5}
\ee
where $\hat{t}_{k\ell}^{\, \dagger}=( n^{-1} \hat{s}_{1,k\ell}  +  m^{-1}  \hat{s}_{2,k\ell}  )^{-1/2} ( \hat \sigma^\dagger_{1,k\ell} - \hat \sigma^\dagger_{2,k\ell})$ with $\hat \sigma^\dagger_{1,k\ell}  =  n^{-1} \sn g_{i}  \{ (X_{ik}-\bar{X}_k)(X_{i\ell}-\bar{X}_\ell)-\hat{\sigma}_{1,k\ell}  \}$ and $
     \hat \sigma^\dagger_{2,k\ell}  = m^{-1} \sm g_{n+j}  \{(Y_{jk}-\bar{Y}_k)(Y_{j\ell}-\bar{Y}_\ell)-\hat{\sigma}_{2,k\ell}  \}.$

(III). The critical value ${c}_{\alpha}$ is defined as the upper $\alpha$-quantile of $\wh{T}^\dagger_{\max}$
conditional on $\{\mathcal{X}_n, \mathcal{Y}_m\}$; that is,
$
     c_{\alpha} = \inf \big\{ t\in \br: \P_g( \wh{T}_{\max}^\dagger >t) \leq \alpha \big\},$
     where $\P_g$ denotes the probability measure induced by the Gaussian random variables $\{g_i\}_{i=1}^N$ with $\mathcal{X}_n$ and $\mathcal{Y}_m$ being fixed.
%\label{eq2.6}

This algorithm combines the ideas of multiplier bootstrap and parametric bootstrap. The principle of parametric bootstrap allows $\hat{t}_{k\ell}^{\,\dagger}$'s constructed in step (II) to retain the {covariance structure of} $\hat{t}_{k\ell}$'s. The validity of multiplier bootstrap is guaranteed by the multiplier central limit theorem, see \cite{VW96} for traditional fixed- and low-dimensional settings and
 \cite{ChernozhukovChetverikovKato_2013} for more recent development in high dimensions.

For implementation, it is natural to compute the critical value ${c}_{\alpha}$ via Monte Carlo simulation by
${c}_{B,\alpha}=\inf \{ t\in \br: 1-\widehat{F}_B(t) \leq \alpha \}$, where $\widehat{F}_B(t)=B^{-1}\sum_{b=1}^BI( \wh T^\dagger_b \leq t)$ and $\wh T_{1}^{\dagger},\ldots,\wh T_{B}^\dagger$ are $B$ independent realizations of $ \wh{T}_{\max}^\dagger$ in (\ref{eq2.5}) by repeating steps (I) and (II). For any prespecified $\alpha \in (0,1)$, the null hypothesis (\ref{covariance.test}) is rejected whenever $\wh{T}_{\max} > {c}_{B,\alpha}$.

The main computational cost of our procedure for {computing} the critical value $c_{B,\alpha}$ only involves generating $N B$ independent and identically distributed $N(0,1)$ variables. It took only 0.0115 seconds to generate one million such realizations based on a computer equipped with Intel(R) Core(MT) i7-4770 CPU $@$ 3.40GHz. Hence even taking $B$ to be in the order of thousands, our procedure can be easily accomplished efficiently when $p$ is large.
\subsection{Theoretical properties}
\label{theory}
The difference between ${c}_{\alpha}$ and its Monte Carlo counterpart ${c}_{B,\alpha}$  is usually negligible for a large value of $B$. In this section, we study the asymptotic properties of the proposed test $\Psi_\alpha=I\{ \wh{T}_{\max}> c_\alpha\}$ under both the null hypothesis \eqref{covariance.test} and a sequence of local alternatives.

For the asymptotic properties, we only require the following relaxed regularity conditions.
Let $K>0$ be a finite constant independent of $n, m$ and $p$.

\begin{itemize}
\item[\textbf{(C1).}] $\{ \mathbb{E} ( |X_{k}|^{2r} )  \}^{1/r} \leq K \sigma_{1,kk}$, $\{  \mathbb{E}( |Y_{k}|^{2r} ) \}^{1/r}  \leq  K \sigma_{2,kk}$ uniformly in $k=1,\ldots, p$, for some $r\geq 4$.

\item[\textbf{(C2).}] $\max_{1\leq k\leq p} \mathbb{E}\{\exp( \kappa X_{ k}^2/\sigma_{1,kk})\} \leq K$ and $\max_{1\leq \ell \leq p}  \mathbb{E}\{\exp(\kappa Y_{ \ell}^2/\sigma_{2,\ell \ell})\} \leq K$ for some $\kappa >0$.

\item[\textbf{(C3).}] $\min_{1\leq k\leq \ell\leq p} {s_{1,k\ell}}/(\sigma_{1,kk} \sigma_{1,\ell \ell}) \geq c$ and $\min_{1\leq k\leq \ell\leq p}  {s_{2,k\ell}}/(\sigma_{2,kk} \sigma_{2,\ell \ell}) \geq c$ for some $c>0$.

\item[\textbf{(C4).}] $n$ and $m$ are comparable, i.e. ${n}/{m}$ is uniformly bounded away from zero and infinity.
\end{itemize}

Assumptions (C1) and (C2) specify the polynomial-type and exponential-type tails conditions on the underlying distributions of $\bX$ and $\bY$, respectively. Assumption (C3) ensures that the random variables $\{ (X_{ k}-\mu_{1k} )(X_{ \ell} - \mu_{1\ell }) \}_{1\leq k, \ell  \leq p}$ and $\{ (Y_{ k} - \mu_{2k})(Y_{ \ell} - \mu_{2\ell }) \}_{1\leq k, \ell  \leq p}$ are non-degenerate.
The moment assumptions, (C1)--(C3), for the proposed procedure are similar to Conditions (C2) and  (C2$^*$) in \cite{CaiLiuXia_2013}. Assumption (C4) is a standard condition in two-sample hypothesis testing problems. As discussed before, no structural assumptions on the unknown covariances are imposed for the proposed procedure. Theorem \ref{asymptotic.size} below shows that, under these mild moment and regularity conditions, the proposed test $\Psi_\alpha$ with $c_\alpha$ defined in Section \ref{sec22a} has an asymptotically $\alpha$.

\begin{theorem} \label{asymptotic.size}
Suppose that Assumptions (C3) and (C4) hold. If  either Assumption (C1) holds with
$p = O(  n^{r/2-1-\de} )$ for some constant $\delta>0$ or Assumption (C2) holds with $\log p = o( n^{1/7} )$,  then as $n,m\rightarrow \infty$, $\P_{H_0} (\Psi_\alpha=1  )  \rightarrow \alpha$ uniformly over $\alpha \in (0,1)$.
\end{theorem}

\begin{remark} The asymptotic validity of the proposed test is obtained
without imposing structural assumptions on $\bSigma_1$ and $\bSigma_2$, nor do we specify any a priori parametric shape constraints of the data distributions, such as Condition~A3 in \cite{LiChen_2012} or Conditions~(C1) and (C3) in \cite{CaiLiuXia_2013}.
\end{remark}

Next, we investigate the asymptotic power of $\Psi_\alpha$. It is known that the $L_\infty$-type test statistics are preferred to the $L_2$-type statistics, including those proposed by  \cite{Schott_2007} and \cite{LiChen_2012}, when sparse alternatives are under consideration. As discussed in Section \ref{intro}, the scenario in which the difference between $\bSigma_1$ and $\bSigma_2$ occurs only at a small number of locations is of great interest in a variety of scientific studies. Therefore, we focus on the local sparse alternatives  characterized by the following class of matrices
\begin{align}
   \mathcal{M}(\gamma)  &  = \bigg\{   (\bSigma_1, \bSigma_2): \,
  \bSigma_1 \mbox{ and } \bSigma_2  \mbox{ are  positive semi-definite matrices satisfying } \nn \\
  & \qquad \qquad\qquad\quad  \mbox{Assumption (C3) and} \max_{1\leq k\leq \ell\leq p} \frac{|\sigma_{1,k\ell}-\sigma_{2,k\ell}|}{ ( n^{-1} s_{1,k\ell} + m^{-1} s_{2,k\ell}   )^{1/2} } \geq  (\log p)^{1/2}  \gamma  \bigg\}. \nn %\label{local.alternatives}
\end{align}
Theorem \ref{power.consistency} below shows that, with probability tending to 1, the proposed test $\Psi_\alpha$ is able to distinguish $H_0$ from the alternative $H_1$ whenever $(\bSigma_1, \bSigma_2) \in \mathcal{M}(\gamma)$ for some $\gamma >2$.

\begin{theorem}  \label{power.consistency}

Suppose that Assumptions (C3) and (C4) hold. If  either Assumption 1 holds with
$p = O(  n^{r/2-1-\de} )$ for some constant $\delta>0$ or Assumption 2 holds with $\log p = o( n^{1/2} )$, then as $n, m\rightarrow\infty$,  $ \inf_{(\bSigma_1, \bSigma_2) \in \mathcal{M}(\gamma)} \P_{H_1}  ( \Psi_\alpha =1  ) \rightarrow 1$ for any $\gamma>2$.
\end{theorem}

Theorem 2 of \cite{CaiLiuXia_2013} requires $\gamma=4$ to guarantee the consistency of their procedure. Moreover, they showed that the rate $(\log p)^{1/2} n^{-1/2}$ for the lower bound of the maximum magnitude of the entries of $\bSigma_1-\bSigma_2$ is minimax optimal, that is, for any $\alpha, \beta>0$ satisfying $\alpha+\beta<1$, there exists a constant $\gamma_0>0$ such that
$
    \inf_{(\bSigma_1,\bSigma_2) \in \mathcal{M}(\gamma_0)} \sup_{T_\alpha \in \mathcal{T}_\alpha} \P_{H_1} ( T_\alpha =1  ) \leq 1-\beta
$
for all sufficiently large $n$ and $p$, where $\mathcal{T}_\alpha$ is the set of $\alpha$-level tests over the collection of distributions satisfying Assumptions (C1) and (C2). Hence, our proposed test also enjoys the optimal rate and is powerful against sparse~alternatives.

\vspace{-10pt}

\section{Simulation studies}
\label{simulation.sec}

In this section, we compare the finite-sample performance of the proposed new test with that of several alternative testing procedures, including \cite{Schott_2007} (Sc hereafter), \cite{LiChen_2012} (LC hereafter) and \cite{CaiLiuXia_2013} (CLX hereafter). We generated two independent random samples $\{\bX_i\}_{i=1}^n$ and $\{\bY_{j}\}_{j=1}^m$ such that  $\bX_i=\bSigma_{1,*}^{1/2}\bZ_i^{(1)}$ and $\bY_j=\bSigma_{2,*}^{1/2}\bZ_j^{(2)}$ with $\bZ_i^{(1)}=(Z_{i1}^{(1)},\ldots,Z_{ip}^{(1)})^\T$ and $\bZ_j^{(2)}=(Z_{j1}^{(2)},\ldots,Z_{jp}^{(2)})^\T$, where $Z_{i1}^{(1)},\ldots,Z_{ip}^{(1)}$ and $Z_{j1}^{(2)},\ldots,Z_{jp}^{(2)}$ are two sets of independent and identically distributed (i.i.d.) random variables with variances $\sigma_{Z,1}^{2}$ and $\sigma_{Z,2}^2$, such that $\bSigma_1=\sigma_{Z,1}^2\bSigma_{1,*}$ and $\bSigma_2=\sigma_{Z,2}^2\bSigma_{2,*}$. We assess the performance of the aforementioned tests under the null hypothesis \eqref{covariance.test}. Let $\bSigma_{1,*}=\bSigma_{2,*}=\bSigma_*$ and consider the following four different covariance structures for $\bSigma_*$.

\vspace{-.255cm}
\begin{itemize}
\setlength\itemsep{2.5pt}
\item M1 (Block diagonals): Set $\bSigma_* = \bD^{1/2}\bA\bD^{1/2}$, where $\bD$ is a diagonal matrix whose diagonals are i.i.d. random variables drawn from $\mbox{Unif}(0.5,2.5)$. Let $\bA=(a_{k \ell})_{1\leq k, \ell \leq p}$, where $a_{kk}=1$, $a_{k \ell}=0.55$ for $10(q-1)+1\leq k \neq \ell \leq 10q$ for $q=1,\ldots, \lfloor p/10 \rfloor$, and $a_{k \ell}=0$ otherwise.

\item M2 (Slow exponential decay): Set $\bSigma_* = (\sigma_{k \ell,*})_{1\leq k, \ell \leq p} $, where $\sigma_{k \ell,* } = 0.99^{|k - \ell |^{1/3}}$.

\item M3 (Long range dependence): Let $\bSigma_* =(\sigma_{k \ell,* })_{1\leq k , \ell \leq p}$ with i.i.d. $\sigma_{kk,*}\sim \mbox{Unif}(1,2)$, and $\sigma_{k \ell,* } =\rho_\alpha(|k -\ell |)$, where
$\rho_\alpha(d)= \{(d+1)^{2H}+(d-1)^{2H}-2d^{2H}\}/2$ with $H=0.85$.

\item M4 (Non-sparsity): Define matrices $\bF=(f_{k \ell})_{1\leq k , \ell \leq p}$ with $f_{kk}=1,f_{k,k+1}=f_{\ell + 1 , \ell} = 0.5$, $\bU \sim \mathcal{U}(\mathcal{V}_{p,k_0})$, the uniform distribution on the Stiefel manifold (i.e. $\bU\in \mathbb{R}^{p\times k_0}$ and $\bU^{\T}\bU=\bI_{k_0}$, the $k_0$-dimensional identity matrix), and diagonal matrix $\bD$ with diagonal entries being i.i.d. $\mbox{Unif}(1,6)$ random variables. We took $k_0=10$ and $\bSigma_*=\bD^{1/2}(\bF+\bU \bU^{\T})\bD^{1/2}$.
\end{itemize}
\vspace{-.255cm} In practice, non-Gaussian measurements are particularly common for high throughput data, such as data with heavy tails in microarray experiments and data of count type with zero-inflation in image processing. To mimic these practical scenarios, we considered the following three models of innovations $Z_{ik}^{(1)}$ and $Z_{jk}^{(2)}$ to generate data. \vspace{-.255cm}
\begin{itemize}
\setlength\itemsep{2.5pt}
\item (D1)  Let $Z_{ik}^{(1)}$ and $Z_{jk}^{(2)}$ be Gamma random variables: $Z_{ik}^{(1)},Z_{jk}^{(2)}\sim \text{Gamma}(4,10)$.

\item (D2) Let $Z_{ik}^{(1)}$ and $Z_{jk}^{(2)}$ be zero-inflated Poisson random variables: $Z_{ik}^{(1)}, Z_{jk}^{(2)}\sim \text{Pois}(1000)$ with probability $0.15$ and equals to zero with probability $0.85$.

\item (D3) Let $Z_{ik}^{(1)}$ and $Z_{jk}^{(2)}$ be Student's $t$ random variables: $Z_{ik}^{(1)}\sim t_5$ and $Z_{jk}^{(2)}\sim t_5(\mu)$ with non-central parameter $\mu$ drawn from $\textrm{Unif}(-2,2)$.
 \end{itemize}
\vspace{-.255cm}
For the numerical experiments, $(n_1,n_2)$ was taken to be $(45,45)$ and $(60,80)$, and the dimension $p$ took value in $\{ 80,280,500,1000\}$. To compute the critical value for the proposed test $\Psi_{B,\alpha}$,  $B$ was taken to be $1500$.

\begin{sidewaystable}[h]

  \caption{Empirical sizes of the proposed test $\Psi_{B,\alpha}$ along with those of the tests by \citet{LiChen_2012} (LC), \citet{Schott_2007} (Sc), and \citet{CaiLiuXia_2013} (CLX) for data generated by data models D1--D3 with covariance structures M1 and M2. Results are based on 1000 replications with $\alpha=0.05$, $(n_1,n_2)=(45,45)$ and $(60,80)$. }
   \centering
          {
    \begin{tabular}{ccccccccccccc}\toprule\toprule
  &  \multicolumn{4}{c}{D1}      &      \multicolumn{4}{c}{D2}    &    \multicolumn{4}{c}{D3}                       \\[1ex]
$p$ &  80&280&500&1000&80&280&500&1000&80&280&500&1000\\[1ex]  \midrule
 & \multicolumn{12}{c}{Covariance structure M1 with $(n_1,n_2)=(45,45)$}     \\[1ex]  \midrule
 $\Psi_{B,\alpha}$  & 0.053&    0.053&  0.053&  0.059&  0.072&  0.072&  0.094   &0.077  &0.032  &0.028  &0.029& 0.032\\[.5ex]
 LC  & 0.066&   0.057&  0.056&  0.059&  0.089&  0.084&  0.073   &0.059  &0.326  &0.325  &0.300& 0.309\\
 [.5ex]
 Sc & 0.119&    0.109&  0.104&  0.115&  0.611&  0.566&  0.616   &0.608  &1.000  &1.000& 1.000&  1.000\\
 [.5ex]
CLX & 0.045&    0.038&  0.027&  0.031&  0.069&  0.062&  0.047   &0.064  &0.015  &0.009& 0.009   &0.007\\[0.5ex] \midrule
  & \multicolumn{12}{c}{Covariance structure M1 with $(n_1,n_2)=(60,80)$}\\[0.5ex] \midrule
 $\Psi_{B,\alpha}$ & 0.038& 0.033&  0.037&  0.032&  0.035&  0.045&  0.050   &0.052  &0.017& 0.029&  0.025   &0.027\\[.5ex]
LC &0.060   &0.065& 0.057&  0.055&  0.042&  0.069&  0.055   &0.059  &0.345& 0.369&  0.361   &0.371\\[.5ex]
 Sc & 0.104 &0.087  &0.111& 0.101   &0.622& 0.641&  0.613   &0.651  &1.000  &1.000  &1.000  &1.000\\[.5ex]
CLX  & 0.036    &0.027  &0.024  &0.026& 0.031   &0.034& 0.046   &0.028  &0.010& 0.013   &0.003& 0.004  \\[0.5ex]  \midrule
& \multicolumn{12}{c}{Covariance structure M2 with $(n_1,n_2)=(45,45)$}     \\[0.5ex]  \midrule
 $\Psi_{B,\alpha}$ & 0.053& 0.057&  0.052&  0.068&  0.051&  0.064&  0.090&  0.090&  0.035   &0.027& 0.032&  0.038\\ [0.5ex]
LC &0.056&  0.068&  0.067&   0.080&  0.096&  0.091&  0.077&  0.088&  0.336   &0.328  &0.348& 0.310\\ [0.5ex]
 Sc &0.076& 0.079&  0.086&  0.089&  0.348&  0.325&  0.166&  0.115&  1.000&  1.000   &1.000  &1.000\\ [0.5ex]
CLX  & 0.054&   0.041&  0.033&  0.037   &0.041& 0.056&  0.049&  0.070   &0.014  &0.007  &0.009& 0.010\\
[0.5ex] \midrule
  & \multicolumn{12}{c}{Covariance structure M2 with $(n_1,n_2)=(60,80)$}\\  [0.5ex] \midrule
 $\Psi_{B,\alpha}$ &0.044&  0.039   &0.032& 0.032   &0.037  &0.033& 0.043&  0.053   &0.020  &0.013  &0.022  &0.028\\[.5ex]
LC & 0.076& 0.090   &0.093  &0.086  &0.086  &0.079  &0.059& 0.091   &0.325  &0.344  &0.338  &0.374\\[.5ex]
Sc &0.118&  0.080   &0.091  &0.078  &0.454  &0.137  &0.342  &0.142  &1.000& 1.000   &1.000  &1.000\\[.5ex]
CLX & 0.040&    0.042   &0.026  &0.027  &0.032  &0.023  &0.034  &0.042  &0.012  &0.005& 0.008   &0.004
  \\ \bottomrule
\end{tabular}}
\label{size12}
\end{sidewaystable}

\begin{sidewaystable}[h]

   \centering
      \caption{Empirical sizes of the proposed test $\Psi_{B,\alpha}$ along with those of the tests by \citet{LiChen_2012} (LC), \citet{Schott_2007} (Sc), and \citet{CaiLiuXia_2013} (CLX) for data generated by data models D1--D3 with covariance structures M3 and M4. Results are based on 1000 replications with $\alpha=0.05$, $(n_1,n_2)=(45,45)$ and $(60,80)$.
      } {
    \begin{tabular}{ccccccccccccc}\toprule\toprule
  &  \multicolumn{4}{c}{D1}      &      \multicolumn{4}{c}{D2}    &    \multicolumn{4}{c}{D3}                       \\  [1ex]
  $~p$ &  80&280&500&1000&80&280&500&1000&80&280&500&1000\\ [1ex] \midrule
& \multicolumn{12}{c}{Covariance structure M3 with $(n_1,n_2)=(45,45)$}     \\  [1ex] \midrule
 $\Psi_{B,\alpha}$ &0.052&  0.062   &0.041  &0.056  &0.065  &0.072  &0.075  &0.081  &0.029  &0.028  &0.033  &0.037\\ [0.5ex]
LC &0.064&  0.067   &0.058  &0.058  &0.101  &0.065  &0.055  &0.054  &0.321  &0.302  &0.311  &0.323\\ [0.5ex]
Sc&0.114&   0.104   &0.108  &0.114  &0.580  &0.611  &0.626  &0.595  &1.000  &1.000  &1.000  &1.000\\ [0.5ex]
CLX&0.041&  0.046   &0.033  &0.033  &0.059  &0.065  &0.042  &0.071  &0.016  &0.010   &0.012  &0.006\\
 [0.5ex] \midrule
  & \multicolumn{12}{c}{Covariance structure M3 with $(n_1,n_2)=(60,80)$}\\ [0.5ex] \midrule
 $\Psi_{B,\alpha}$ &0.039&  0.038   &0.036  &0.040  &0.038  &0.043  &0.043  &0.053  &0.018  &0.023  &0.024  &0.025\\ [0.5ex]
LC& 0.066&  0.063   &0.074  &0.040  &0.086  &0.053  &0.072  &0.068  &0.337  &0.335  &0.343  &0.342\\ [0.5ex]
Sc&0.108&   0.104   &0.134  &0.098  &0.651  &0.674  &0.644  &0.662  &1.000  &1.000  &1.000  &1.000\\ [0.5ex]
CLX&0.034&  0.032   &0.029  &0.031  &0.028  &0.035  &0.035  &0.025  &0.006  &0.011  &0.006  &0.005  \\[0.5ex] \midrule
 & \multicolumn{12}{c}{Covariance structure M4 with $(n_1,n_2)=(45,45)$}     \\ [0.5ex] \midrule
 $\Psi_{B,\alpha}$ &0.054   &0.056  &0.056  &0.078  &0.052  &0.079  &0.086  &0.086  &0.021  &0.031  &0.031  &0.027\\[0.5ex]
LC&0.063    &0.068  &0.055  &0.060  &0.064  &0.070  &0.070  &0.053  &0.323  &0.311  &0.343  &0.318\\[0.5ex]
Sc&0.117    &0.107  &0.098  &0.120  &0.595  &0.606  &0.632  &0.621  &1.000  &1.000  &1.000  &1.000\\[0.5ex]
CLX&0.049   &0.049  &0.043  &0.037  &0.045  &0.066  &0.040  &0.076  &0.009  &0.011  &0.004  &0.004\\[0.5ex] \midrule
  & \multicolumn{12}{c}{Covariance structure M4 with $(n_1,n_2)=(60,80)$}\\[0.5ex] \midrule
 $\Psi_{B,\alpha}$ &
0.044   &0.050  &0.036  &0.042  &0.047  &0.042  &0.047  &0.055  &0.029  &0.013  &0.022  &0.024\\[0.5ex]
LC&0.053    &0.058  &0.054  &0.055  &0.104  &0.049  &0.070  &0.051  &0.340  &0.334  &0.335  &0.341\\[0.5ex]
Sc&0.110    &0.100  &0.117  &0.111  &0.618  &0.650  &0.641  &0.682  &1.000  &1.000  &1.000  &1.000\\[0.5ex]
CLX&0.038   &0.036  &0.036  &0.032  &0.036  &0.037  &0.039  &0.025  &0.016  &0.004  &0.006  &0.004
  \\ \bottomrule
\end{tabular}}
\label{size34}
\end{sidewaystable}

Tables~\ref{size12} and \ref{size34} display the empirical sizes of $\Psi_{B,\alpha}$, the LC test, Sc test and CLX test. For both the Gamma and zero-inflated Poisson data (models D1 and D2), the Sc test fails to maintain the nominal size while the other three tests maintain the significance level reasonably well. For the $t$-distributed data (model D3), both the Sc and LC tests had distorted empirical sizes. In contrast, the proposed test $\Psi_{B,\alpha}$ has empirical size closer to the nominal level for the $t$-distributed data while the CLX test is more conservative. This confirms the early discussions that the limiting distribution based approach for $L_\infty$-type test procedure can sometimes be conservative. Compared to {the} existing {methods}, $\Psi_{B,\alpha}$ has a much wider applicability as it requires no
structural assumptions on the unknown covariances and circumvents the issue of slow convergence of $L_\infty$-type statistic to its limiting distribution. Overall, $\Psi_{B,\alpha}$ maintains the nominal size in finite sample reasonably well and is robust against unknown covariance structures as well as data generation mechanisms.

\begin{figure}[h!]
    \noindent

      \makebox[1\textwidth][c]{  \subfigure[Covariance structure M1]
    {
   \includegraphics[width=0.9\linewidth,height=4.25cm]{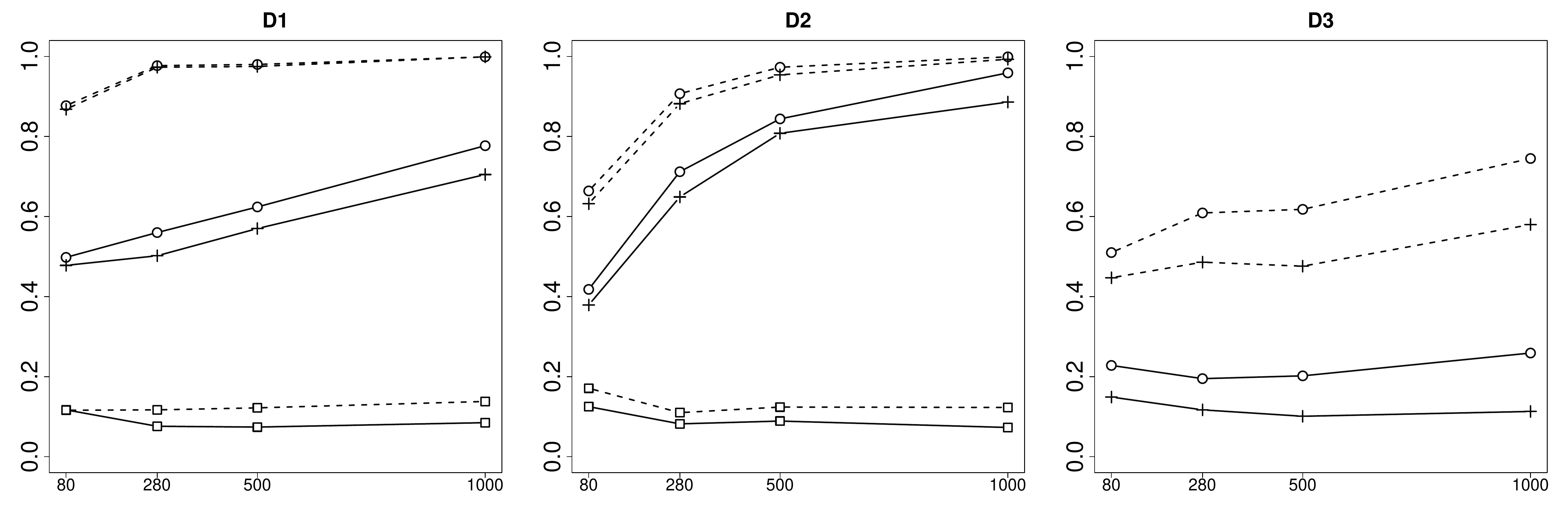}
    }
    }

          \makebox[1\textwidth][c]{  \subfigure[Covariance structure M2]
    {
    \includegraphics[width=0.9\linewidth,height=4.25cm]{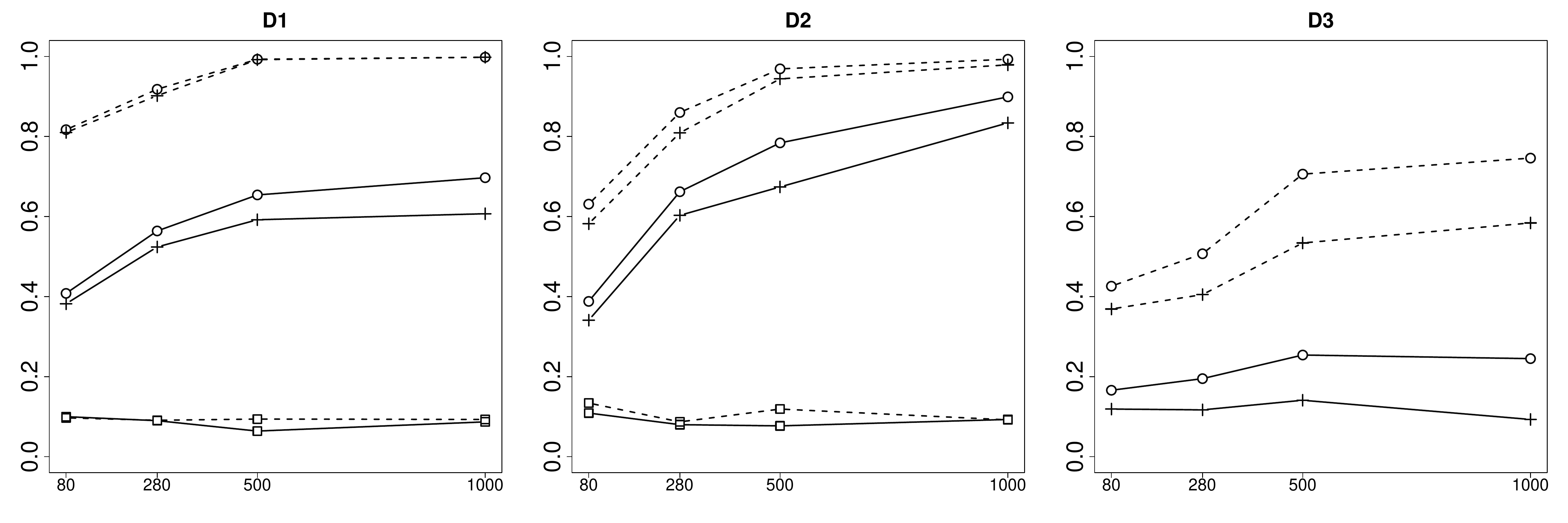}
          }

    }

          \makebox[1\textwidth][c]{  \subfigure[Covariance structure M3]
    {
   \includegraphics[width=0.9\linewidth,height=4.25cm]{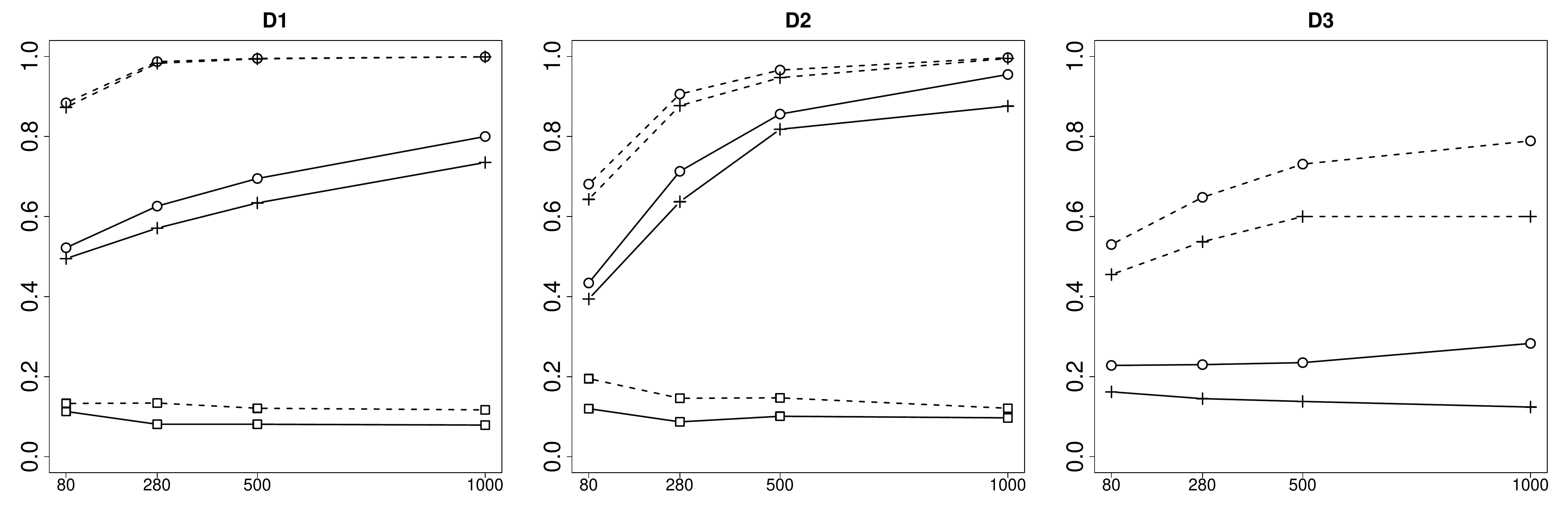}
    }
    }

          \makebox[1\textwidth][c]{  \subfigure[Covariance structure M4]
    {
    \includegraphics[width=0.9\linewidth,height=4.25cm]{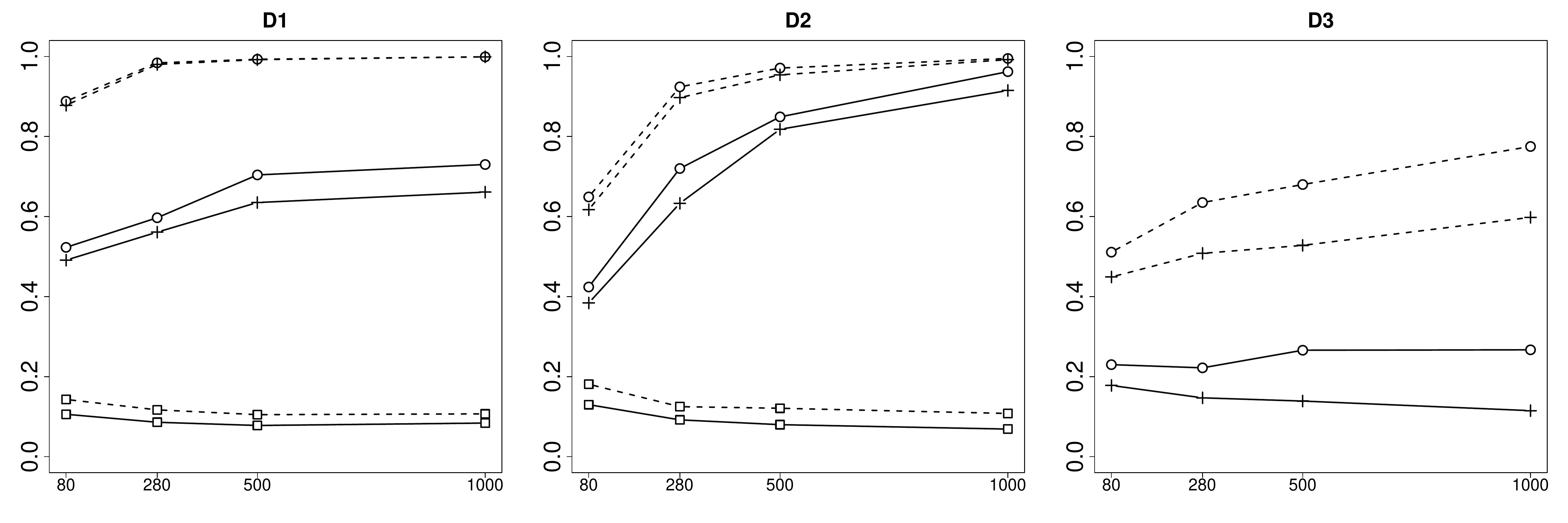}
          }
}
   \caption{Comparison of empirical powers for data generated by data models D1--D3 with different covariance structures. In each panel, horizontal and vertical axes depict dimension $p$ and empirical powers, respectively; and unbroken lines and dashed lines represent the results for $(n_1,n_2)=(45,45)$ and $(60,80)$, respectively. The different symbols on the lines represent different tests experimented in the study, where $\circ$, $\oblong$, and $+$ indicate the proposed test, tests by \citet{LiChen_2012} and \citet{CaiLiuXia_2013}, respectively. Results are based on 1000 replications with $\alpha=0.05$.}
    \label{pr1}
\end{figure}

To evaluate the power performance against relatively sparse alternatives, we define a perturbation matrix $\bQ$ with $\lfloor 0.05 p\rfloor$ random non-zero entries. Half of the non-zero entries are randomly allocated in the upper triangle part of $\bQ$ and the others are in its lower triangle part by symmetry. The magnitudes of non-zero entries are randomly generated from $ \mbox{Unif}(\tau/2,3\tau/2)$ with $\tau=8\max\{\max_{1\leq k\leq p} \sigma_{kk,*}, (\log p)^{1/2} \}$, where $\sigma_{kk,*}$'s are the diagonal entries of $\bSigma_*$ specified before. We take $\bSigma_{1,*}=\bSigma_*+\lambda_0 \bI_p$ and $\bSigma_{2,*}=\bSigma_*+\bQ+\lambda_0 \bI_p$, where $\lambda_0=|\min\{ \lambda_{\text{min}}(\bSigma_*+\bQ),\lambda_{\text{min}}(\bSigma_*)\} |+0.05$ with $\lambda_{\text{min}}(\bA)$ denoting the smallest eigenvalue of matrix $\bA$. For the Gamma and zero-inflated Poisson data (panels for D1 and D2 in Figure~\ref{pr1}), only the proposed test $\Psi_{B,\alpha}$, the LC and CLX tests are considered since the Sc test is no longer applicable due to inflated sizes; and similarly, for the $t$-distributed data (panels for D3 in Figure~\ref{pr1}), only $\Psi_{B,\alpha}$ and the CLX test are considered.

Figure~\ref{pr1} displays empirical power comparisons. We see that the proposed test $\Psi_{B,\alpha}$ and the CLX test are substantially more powerful than the LC test against sparse alternatives for the Gamma and zero-inflated Poisson data (data models D1 and D2) under different covariance structures. As the number of non-zero entries of $\bSigma_1-\bSigma_2$ grows in $p$, both the proposed test $\Psi_{B,\alpha}$ and the CLX test gain powers while the LC test do not gain much due to the sparsity of $\bSigma_1-\bSigma_2$. For the Gamma and zero-inflated Poisson data, the proposed test is slightly more powerful than the CLX test when the sample size is small and the two tests are closely comparable as the sample size increasing. For $t$-distributed data (data model D3), $\Psi_{B,\alpha}$ is more powerful than the CLX test and gains more powers along increasing sample sizes and dimensions. In summary, $\Psi_{B,\alpha}$ outperforms the other three for sparse alternatives of interest. More simulation results are reported in the supplementary material.

\vspace{-10pt}

\section{Application of the proposed procedure in gene clustering}
\label{sec51}

The primary goal of gene clustering is to group genes with similar expression patterns together, which usually provides insights on their biological functions or regulatory pathways. In genomic studies, gene clustering has been employed for detecting co-expression gene sets \citep{Dhaeseleer05,SES2002}, %Y03
 identifying functionally related genes \citep{YST2001}, and discovering large groups of genes suggestive of co-regulation by common factors%\citep{H05}
, among other applications.

Consider a random sample $\mathcal{X}_n=\{\bX_1,\ldots, \bX_n\}$ of $n$ independent observations from $\bX=(X_1,\ldots,X_p)^{\T}$ with covariance $\bSigma_1=(\sigma_{1,k\ell})_{1\leq k,\ell\leq p}$ and correlation $\bR_1=(\rho_{1,k\ell})_{1\leq k,\ell\leq p}$, where $\bX_{i}$ records {the} expression levels of $p$ genes {from} subject $i$. To cluster the genes based on their expression levels, some dissimilarity or proximity measure for the $p$ genes, or equivalently, the $p$ variables, is calculated based on $\mathcal{X}_n$, to which clustering algorithms are applied. Gene clustering can therefore be achieved via clustering the variables. To discover the clustering structure of variables, it is intuitive that variables $X_{k}$ and $X_{\ell}$ will be clustered in the same group if $|\rho_{1,k\ell}|$ is large and separated otherwise \citep{WL2008}.
Specifically,
%Variable clustering can be described via the covariance matrix $\Sigma_1$ in the sense that
if there are some clustering structures among variables, then there exists a partition of $\{1,\ldots, p\}$ upon potential permutations, denoted by $\{B_t\}_{t=1}^m$ for some $1\leq m\leq p$, such that $\min_{k,\ell\in B_t}|\rho_{1,k\ell}|>c_1$, and for any $1\leq t \neq t' \leq m$, $\max_{k\in B_t, \ell\in B_{t'}}|\rho_{1,k\ell}|<c_2$, where $c_1, c_2>0$ are positive constants. {The problem is then} closely related to testing one-sample hypotheses that for a given $\Lambda\subseteq \mathcal{I}_p=\{(1,1),\ldots, (1,p), (2,1), \ldots, (2,p),\ldots, (p,p)\}$,  $ H_0^{\Lambda}: \rho_{1,k\ell}=0 ~ \text{for any}~ (k,\ell)\in \Lambda$ versus $H_1^{\Lambda}:  \rho_{1,k\ell} \neq 0 ~ \text{for some}~ (k,\ell)\in \Lambda$, which is equivalent to
\be H_0^{\Lambda}: {\sigma_{1,k\ell}}=0 ~ \text{for any}~ (k,\ell)\in \Lambda \quad  \mbox{versus}  \quad  H_1^{\Lambda}: {\sigma_{1,k\ell}} \neq 0 ~ \text{for some}~ (k,\ell)\in \Lambda.\label{hp1} \ee Testing the hypothesis \eqref{hp1} facilitates recovering the dissimilarity patterns among variables; that is, failing to reject $H_0^{\Lambda}$ indicates the segregation between $X_k$ and $X_\ell$ whenever $(k,\ell) \in \Lambda$.

Motivated by the block-wise estimation method {of} \cite{CS07}, we define $\Lambda$ in the following way. First, we place the covariance matrix $\bSigma_1$ on a $p\times p$ grid indexed by $\mathcal{I}_p$ and partition it with blocks of moderate size. Due to symmetry, we only focus on the upper triangle part. Second, we construct blocks of size $s_0\times s_0$ along the diagonal and note that the last block may be of a smaller size if $s_0$ is not a divisor of $p$. % (in fact, only the upper half of these blocks will be used and we have $\lceil p/s_0 \rceil$ triangles).
Next, we create new blocks of size $s_0\times s_0$ successively toward the top right corner. Similarly as before, blocks to the most right may be of smaller size. The grid, or equivalently, the index set $\mathcal{I}_p$, is partitioned into $S=\lceil p/s_0\rceil(\lceil p/s_0\rceil+1)/2$ sub-regions and we denote by $\Lambda_1,\ldots, \Lambda_S$ the partition of the upper triangle indices $\{(k,\ell): 1\leq k<\ell \leq p\}$.

On each of the sub-regions, we modify the proposed procedure for testing local hypotheses $H_{0}^{\Lambda_s}: \sigma_{1,k\ell}=0$ for any $(k,\ell)\in\Lambda_s$ versus $H_{1}^{\Lambda_s}: \sigma_{1,k \ell} \neq 0$ for some $(k,\ell)\in \Lambda_s$, $s=1,\ldots,S$. We then apply the Benjamini-Hochberg (BH) procedure to control the false discovery rate (FDR) for simultaneously testing $S$ hypotheses. For each $s$, failing to reject the null $H_{0}^{\Lambda_s}$ indicates a segregation between $X_k$ and $X_\ell$ for $(k,\ell)\in \Lambda_s$ and zero will be assigned as the similarity between $X_k$ and $X_\ell$. We summarize this procedure as follows.
%\begin{algorithm*}
%\caption{Computation of the ${\Psi}_\alpha$-driven dissimilarity}  \label{al2} \ \
%\begin{enumerate}
%\item[{\rm(i)}]

(I) Compute the sample covariance matrix $\widehat{\bSigma}_1=(\hat \sigma_{1,k\ell})_{1\leq k,\ell\leq p}$ and $\widehat \bT=(\tilde t_{k\ell})_{1\leq k,\ell\leq p}$, where $\tilde{t}_{k\ell}= n^{1/2} \hat{s}_{1,k\ell}^{-1/2} \hat{\sigma}_{1,k\ell} $ for $\hat s_{1,k\ell}$ defined in
Section \eqref{sec22a}.
%\eqref{cat}.

%\item[{\rm(ii)}]
(II) Independent of $\mathcal{X}_n$,  simulate a sample of size $B$, where for each $b=1,\ldots, B$ and $1\leq k\leq \ell \leq p$, compute $
\tilde{t}^{\, \dagger}_{b,k\ell}   = ( n^{-1} \hat{s}_{1,k \ell} )^{-1/2}  \sn g_{b,i}  \{  (X_{ik}-\bar{X}_k)(X_{i\ell}-\bar{X}_\ell)-\hat{\sigma}_{1,k\ell}  \} ,
$  where $\{g_{b,1},\ldots, g_{b,n}\}$ is a sequence of i.i.d. standard normal random variables.
%\item[{\rm(iii)}]

(III) Partition the $p\times p$ grid as discussed before by $S$ blocks. For each block with entries indexed by $\Lambda_s\subset \mathcal{I}_p$, compute the approximated $p$-value as
$
    \hat p_{s} = 1-\widehat{F}_B \left(\max_{(k,\ell)\in \Lambda_s}\tilde t_{k\ell}\right)
     , \label{bpvalue}
$
where $\widehat{F}_B$ denotes the empirical (conditional) distribution function of $\max_{(k,\ell)\in \Lambda_s} \tilde t_{k\ell}$ given $\mathcal{X}_n$ using the simulated samples $\{ \max_{(k,\ell) \in \Lambda_s} \tilde{t}^{\, \dagger}_{b,k \ell}\}_{b=1}^B$.
%item[{\rm(iv)}]

(IV) Estimate the $q$-values for $\{\hat{p}_s\}_{s=1}^S$ using the BH procedure, denoted by $\{\hat{q}_s\}$. For a prespecified cut-off $\pi$, define the dissimilarity measure by
\be d_{k\ell}= 1-\frac{\tilde{t}_{k\ell}  I (\hat{q}_s <\pi ) }{\max\{\max_{(k,\ell)\in \Lambda_s} \tilde{t}_{k\ell},1\}}~~~\textrm{for any}~~(k,\ell)\in \Lambda_s. \label{dis} \ee
%\end{enumerate}

%\end{algorithm*}

Based on the measure in \eqref{dis}, we can apply clustering algorithms such as the hierarchical clustering for clustering variables and obtain gene clustering. To specify the blocks, we propose the following data-driven selection of $s_0$. The $S$ local hypotheses to be tested simultaneously admit unknown complex dependencies so that the FDR, controlled by the BH procedure, satisfies the general upper bound $ {\rm FDR}\leq (\pi  S_0\log S)/S$ where $S_0$ denotes the number of true null local hypotheses \citep{BY2001}. To control the FDR at the nominal level $\pi$, we need $S \geq S_0 \log S$ which is automatically satisfied when $S=1$ or $s_0$ is large. Therefore, we define a data-driven $s_0$ by $ s_0=\max\{ \lceil \log p \rceil , \min(s:\widehat{S}_0(s)\leq S(s)  [ \log\{ S(s)\} ]^{-1} )\}$, where $S(s)=\lceil p/s \rceil(\lceil p/s \rceil+1)/2$ and $\widehat{S}_0$ is an estimate for the number of true null local hypotheses. In practice, we may also reorder the variables first using methods such as the Isoband algorithm by \cite{WL2008}. A demonstration of the proposed clustering algorithm, as well as comparisons of $d_{k\ell}$ with traditional dissimilarity measures based on the human asthma data, is displayed in the Supplementary Materials.

%\begin{figure}[h!]
%    \centering
 %    \includegraphics[width=0.5\linewidth]{Simulations/realdata/f0b}
 %   \caption{Comparisons of the empirical powers of tests for model M4 and three data models, $(n_1,n_2)=(45,45)$ and $(60,80)$. 1000 replications.}
 %   \label{pr4}
%\end{figure}

\vspace{-10pt}

%=================================================================================================%
\section{Application to analysis of human asthma data}  \label{real}

\subsection{Background}

As a common chronic inflammatory disease of the airways, asthma is caused by a combination of complex genetic and environmental interactions and affects more than 200 million people worldwide as of 2013
 as shown in 2013 World Health  Organization Fact Sheet No. 307.
%and approximately $300,000$ die per year as shown in 2013 World Health  Organization Fact Sheet No. 307.
%Global rates of asthma have increased significantly since 1970s and has been recognized as a major public health issue.
The mechanism and regulatory pathways remain unclear.
%, which influences the effectiveness of clinical treatments \citep{CSB2007}.
We illustrate the proposed new procedures using the human asthma data from the microarray experiment reported by \cite{V2014},
which was aimed to understand the regulatory pathway and mechanism for high nitrative stress, a major characteristic of human severe asthma.
%Recently, to understand the regulatory pathway and mechanism for high nitrative stress, a major characteristic of human severe asthma, microarray %experiments were conducted to compare 3-nitrotyrosines (3NTs), widely recognized proteins with altered functions in the disease, with two major %CD4+ T cell immunity, the Th1 and Th2 pathways \citep{V2014}.
\cite{V2014} identified several novel pathways, %that extend the understanding of genetic basis of asthma. In particular,
including discovering that the Th1 cytokine, IFN-$\gamma$, along or with Th2 regulations, are critical immune agents for the disease development by amplifying epithelial NAD/NADPH thyroid oxidase expression and aiding the production of nitrite. %They also identified the thyroid peroxidase (TPO) as a critical catalyst in 3NT generation that may contribute to the pathology of the disease.

The original microarray gene expression data are available at the NCBI's Gene Expression Omnibus database %\url{http://www.ncbi.nlm.nih.gov/sites/GDSbrowser}
with the Gene Expression Omnibus Series accession number GSE43696. The data consist of %108 samples, including 20 health samples, 50 moderate {asthmatics} and 38 {severe asthmatics} with similar demographic characteristics. We compared the asthma cases with health samples. The health sample consists of
$n_1=20$ health samples and $n_2=88$ patients suffering from moderate or severe asthmatics. We focused on identifying disease-associated GO terms. %defined within the gene ontology (GO) framework. The gene-sets are technically defined in the GO system via structured vocabularies which  produce unique name for a gene-set. %\citep{A00}.
After preliminary filtering steps using the approach in \cite{G2005} and removing genes without appropriate annotations, there remained $24,520$ genes. We excluded GO terms with missing information or   less than 10 genes. {There retained $3,290$ GO terms from the original dataset whose sizes vary from 11 to $8,070$ genes.} For $g=1,\ldots, G$ with $G=3,290$, denote by $\bmu_{h,g}$ and $\bmu_{a,g}$ the mean gene expression levels, and $\bSigma_{h,g}$ and $\bSigma_{a,g}$ the covariance matrices for the
$g^{\text{th}}$ GO term in the control and disease groups, respectively.

\vspace{-10pt}

\subsection{Differential expression analysis}
A commonly used method in differential analysis is the mean-based test that selects interesting GO terms by testing the null hypothesis that overall gene expressions within a GO term are similar across populations \citep{ChenQin_2010,ChangZhouZhou_2014,WPL2015}. Though the mean-based procedure has been successful in detecting differential expressed genes based on the changes in the expression level, recent developments in genomic analysis have revealed the importance to detect genes with changing relationships with other genes in different biological states, and particularly GO terms that change the dependence structures across populations \citep{delaFuente2010}. The discovery of those GO terms with altered dependence structures provides information on critical gene regulation pathways. Consider all the GO terms, we applied the proposed method $\Psi_{B,\alpha}$ to test the global hypotheses \be
 H^c_{0g}: \bSigma_{h,g}=\bSigma_{a,g} \quad \mbox{versus} \quad  H^c_{1g}:\bSigma_{h,g}\neq \bSigma_{a,g}. \label{h2}
\ee
For a comparison, we also applied the LC and CLX tests.

Here, $B=5,000$ Monte Carlo replications were employed to compute the $p$-values for $\Psi_{B,\alpha}$. By controlling the FDR at $2.5\%$ \citep{BY2001}, the proposed test $\Psi_{B,\alpha}$ declared 969 GO terms significant while the LC and CLX tests declared 290 and 524 GO terms significant, respectively. The proposed test $\Psi_{B,\alpha}$ has found more significant GO terms and is less conservative than the others, which is also reflected by the histograms of $p$-values for the three tests displayed in the Supplementary Material. Table \ref{goa1} displays the top 15 most significant GO terms declared by $\Psi_{B,\alpha}$ and also highlights those GO terms that were not detected by the LC and CLX tests. For example, GO:0005887 (integral to plasma membrane) is functionally relevant to the dual oxidases (DUOX2)-thyroid peroxidase interaction and is
important to the mechanism of asthma development \citep{V2014}. %Fortunato10
It is worth noticing that $\Psi_{B,\alpha}$ is able to discover this biologically important GO term that is missed by the others. This further highlights the good performance of our proposed test.
 \begin{table}[h!]
    \centering
    \caption{Top 15 most significant GO terms detected by $\Psi_{B,\alpha}$ with FDR controlled at $2.5\%$, $\flat$ and $\dag$ refer to the GO terms not being declared significant by the CLX test and the LC test, respectively.
      } {
\begin{tabular}{ll}
\toprule
    GO ID & GO term name \\ \midrule
GO:0006886  &intracellular protein transport  $^{\dag}$\\[0.25ex]
GO:0008565  &protein transporter activity $^{\dag}$\\[0.25ex]
GO:0030117  &membrane coat $^{\dag}$\\[0.25ex]
GO:0005515  &protein binding$^{\flat,\dag}$\\[0.25ex]
GO:0016032  &viral reproduction$^{\flat,\dag}$\\[0.25ex]
GO:0005829  &cytosol$^{\dag}$\\[0.25ex]
GO:0000278  &mitotic cell cycle$^{\dag}$\\[0.25ex]
GO:0006334  &nucleosome assembly$^{\dag}$\\[0.25ex]
GO:0034080  &CenH3-containing nucleosome assembly at centromere\\[0.25ex]
GO:0006974  &response to DNA damage stimulus$^{\dag}$\\[0.25ex]
GO:0016874  &ligase activity$^{\dag}$\\[0.25ex]
GO:0032007  &negative regulation of TOR signaling cascade$^{\dag}$\\[0.25ex]
GO:0005887  &integral to plasma membrane$^{\flat,\dag}$\\[0.25ex]
GO:0006997  &nucleus organization$^{\dag}$\\[0.25ex]
GO:0030154  &cell differentiation$^{\dag}$
   \\ \bottomrule
\end{tabular}}
\label{goa1}
\end{table}

In addition, we compared the study on changing intergene relationships across biological states with the traditional differential analysis based on mean expression levels. The proposed test on intergene relationships discovered 268 significant GO terms that were missed by the traditional differential analysis. This reflects the lately growing demands on analyzing gene dependence structures. More details on this comparison are retained in the supplement.

\vspace{-10pt}

\subsection{Gene clustering study on GO terms of interest}
\label{sec52}

\cite{V2014} revealed a novel pathway involving epithelial iNOS, dual oxidases, TPO and the cytokine INF-$\gamma$ to understand the mechanism of human asthma. Multiple transcripts, together with their variants, are related, while their co-regulation mechanisms are less clear. The proposed gene clustering algorithm provides a way to study gene interactions.

For illustration, we focus on the GO terms that were declared significant via testing \eqref{h2} and are related to IFN-$\gamma$ or TPO, and apply our clustering procedure to the sample from the health and disease groups separately to study how the gene clustering alters across two populations. For IFN-$\gamma$, we consider the GO terms 0032689 (negative regulation of IFN-$\gamma$ production), 0060333 (IFN-$\gamma$-mediated signaling pathway) and 0071346 (cellular response to IFN-$\gamma$). For TPO,  the GO terms have been considered include 0004601 (peroxidase activity), 0042446 (hormone biosynthetic process), 0035162 (embryonic hemopoiesis), 0006979 (response to oxidative stress), and 0009986 (cell surface). Their sizes vary from 17 to 439.

\begin{figure}[h!]
    \centering

     \makebox[1\textwidth][c]{  \subfigure[GO:0071346, cellular response to INF-$\gamma$]
    {
  \includegraphics[scale=0.45]{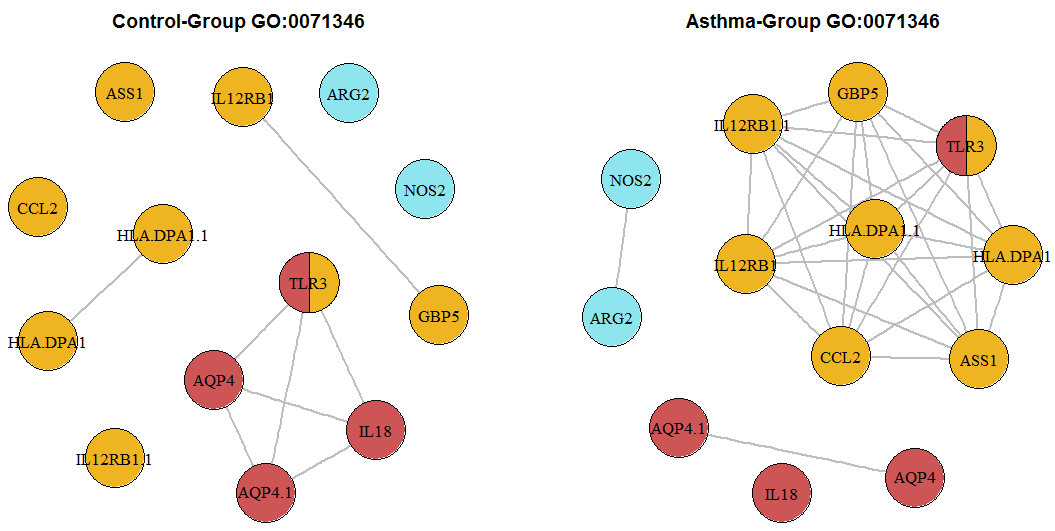}
    }
    }

        \makebox[1\textwidth][c]{  \subfigure[GO:0060333, INF-$\gamma$-mediated signaling pathway]
    {
  \includegraphics[scale=0.45]{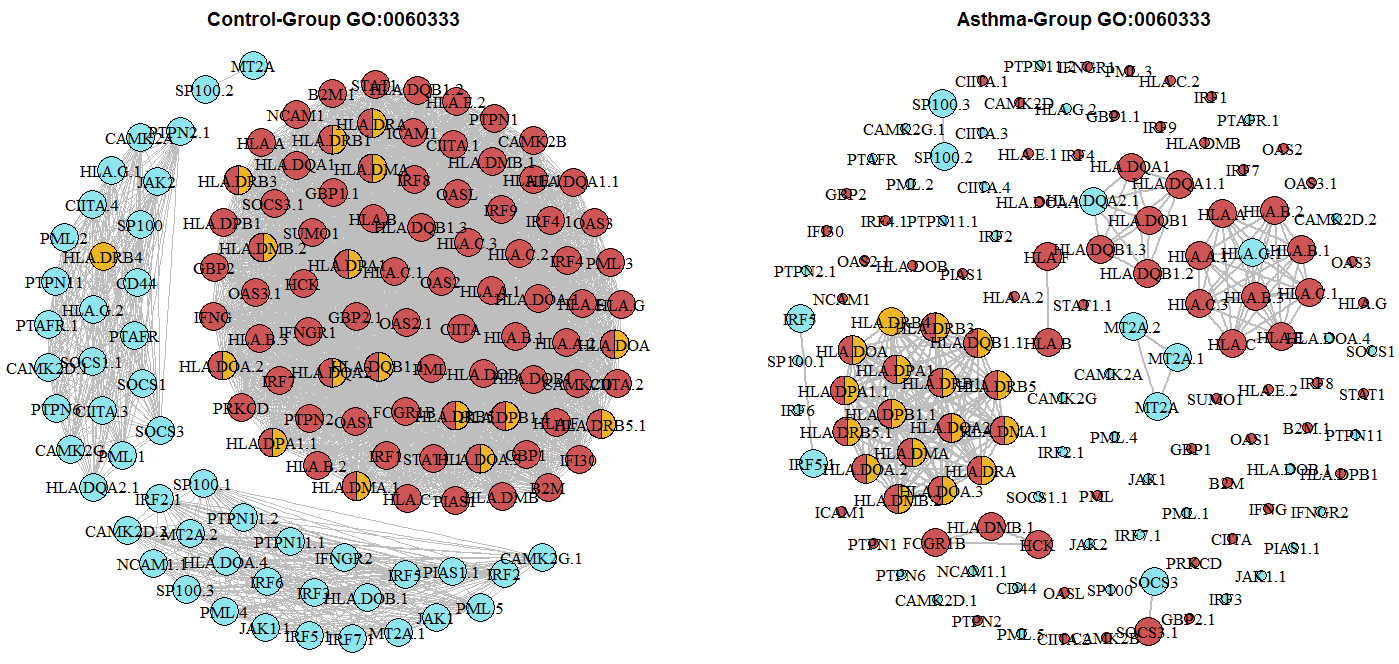}
    }
    }

   \caption{Comparison of clustering structures of GO:0071346, cellular response to INF-$\gamma$ and GO:0060333, INF-$\gamma$-mediated signaling pathway, between health and disease groups using the proposed gene clustering procedure. This figure appears in color in the electronic version of this article.}
    \label{go1}
\end{figure}

\begin{figure}[h!]
    \centering

     \makebox[1\textwidth][c]{  \subfigure[GO:0004601, peroxidase activity]
    {
  \includegraphics[scale=0.45]{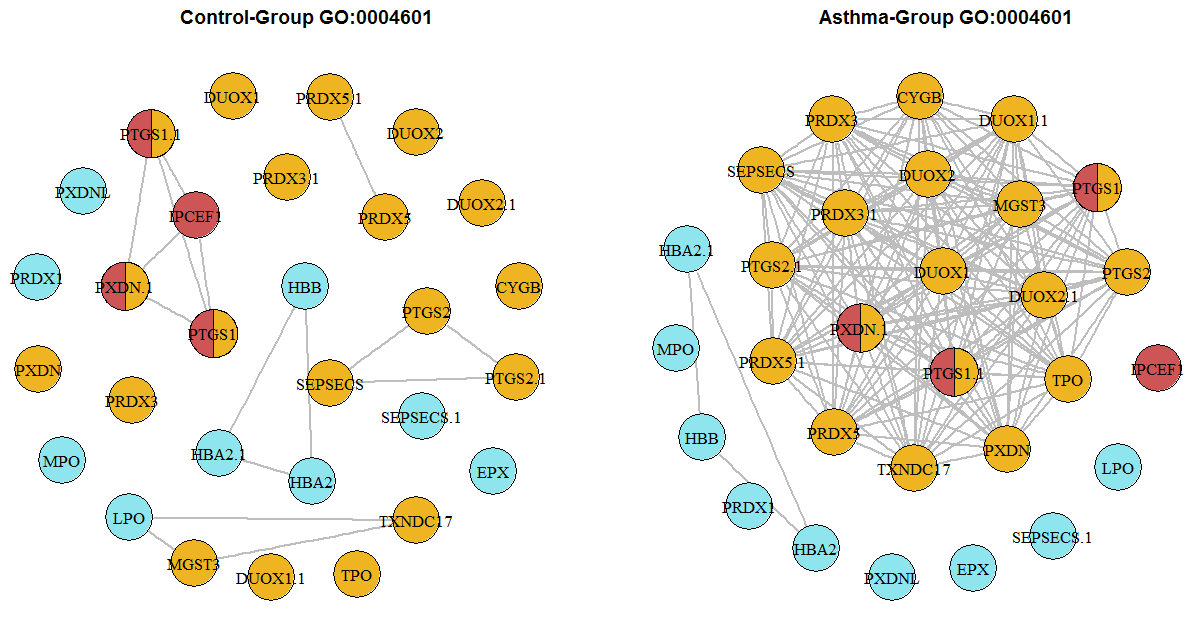}
    }
    }

        \makebox[1\textwidth][c]{  \subfigure[GO:0035162, embryonic hemopoiesis]
    {
 \includegraphics[scale=0.45]{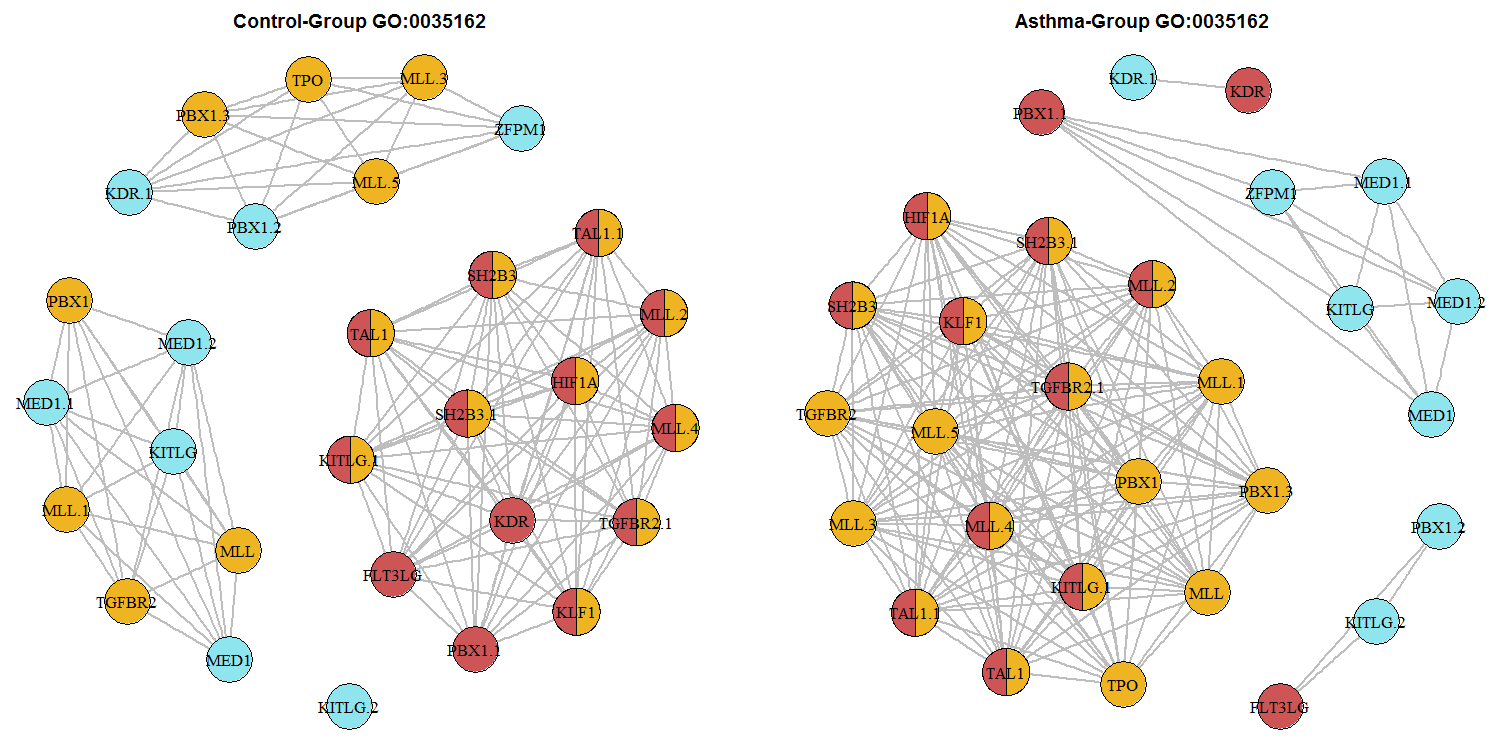}
    }
    }

 \caption{Comparison of clustering structures of GO:0004601, peroxidase activity and GO:0035162, embryonic hemopoiesis, between health and disease groups using the proposed gene clustering procedure. This figure appears in color in the electronic version of this article.}
    \label{go5}
\end{figure}

We take $B =5,000$, $\alpha =0.05$ and use hierarchical clustering algorithm with average linkage. The $\widehat{S}_0$ is estimated using the censored Beta-Uniform mixture model by \cite{ML(2010)} for selecting block size $s_0$. Figures \ref{go1}--\ref{go5} display comparisons of gene clustering between the health and disease groups (more comparisons are included in the Supplementary Material). Each vertex in the figures represents a gene or its variant and is labelled by the corresponding ID. Vertexes connected by edges in gray are clustered into one group, and vertexes in red and yellow belong respectively to the maximum clique in the health and disease groups. Vertexes in both colors belong to the maximum cliques for both groups.

%\begin{figure}[h!]
%    \centering
%  {  \includegraphics[scale=0.4]{Simulations/realdata/go1655_infg} }
%   \caption{Comparison of clustering structures of GO:0060333, INF-$\gamma$-mediated signaling pathway, between health and disease groups using the proposed gene clustering procedure.  }
%    \label{go2}
%\end{figure}

From Figure \ref{go1} we see that for GO:0071346, regarding the cellular response to INF-$\gamma$, genes tend to function more in clusters in the asthma group than those in the health group. Gene TLR3 actively appears in the largest gene clusters for both the health and asthma groups, while gene IL18 is isolated in the asthma group. Gene NOS2 is involved in asthma by co-regulating with ARG2. These suggest that these four genes are important signatures for understanding the effect of INF-$\gamma$ on the asthma progression. Regarding the INF-$\gamma$-mediated signaling pathway, Figure \ref{go1} also shows that compared to the health group, genes seem to preferentially function separately in the asthma group. The original dominating gene clusters are broken into small groups in the presence of the disease.
The different configurations in primary gene clusters between the health and asthma groups for GO:0060333 provide further information on how INF-$\gamma$ influences the iNOS pathway. For the critical enzyme TPO, Figure \ref{go5} shows that genes also tend to function in clusters in the disease group. In the presence of asthma, the gene cluster HBB-HBA2.1-HBA2 is preserved and the gene IPCEF1 is isolated from the original largest gene cluster for GO:0004601. It is interesting to notice that the DUOX2 genes are isolated in the health group but do interact with many genes, particularly with TPO, in the presence of asthma as documented in \cite{V2014}. The identified DUOX2 gene cluster provides a candidate pathway to understand how TPO catalyzes the iNOS-DUOX2-thyroid peroxidase pathway discovered by \cite{V2014}. Last but not least, it can be seen from Figure \ref{go5} that the overall co-regulation patterns remain similar across populations, while those of TPO alters in the presence of asthma.

In summary, based on the proposed procedure, not only can we test the difference in gene dependence, we can also discover the disparity in gene clustering, which reflects the difference in gene clustering patterns between the health and disease groups.

%\begin{figure}[h!]
%    \centering
%  {  \includegraphics[scale=0.4]{Simulations/realdata/go5460_tpo}  }
%   \caption{Comparison of clustering structures of GO:0035162, embryonic hemopoiesis, between health and disease groups using the proposed gene clustering procedure. }
%    \label{go8}
%\end{figure}

\vspace{-10pt}
\section{Conclusion and discussion}\label{discuss}

In this paper, we proposed a computationally fast and effective procedure for testing the equality of two large covariance matrices. The proposed procedure is powerful against sparse alternatives corresponding to the situation where the two covariance matrices differ only in a small fraction of entries. Compared to existing tests, the proposed procedure requires no structural assumptions on the unknown covariance matrices and remains valid under mild conditions. These appealing features grant the proposed test a vast applicability, particularly for real problems arising in genomics. %where unknown and complex structures that involve strong dependence are ubiquitous.
As an important application, we introduced a gene clustering algorithm that enjoys the same nice feature of avoiding imposing structural assumptions on the unknown covariance matrices.

Another interesting and related problem is testing the equality of two precision matrices, which was recently studied by \cite{XTT2015}. In the literature of graphical models, it is common to impose the Gaussian assumption on data so that the conditional dependency can be inferred based on the precision matrix. When the discrepancy between two precision matrices is believed to be sparse, the data-dependent procedure considered in this paper can be extended to comparing them by utilizing the similar $L_\infty$-type statistic discussed in \cite{XTT2015}. It is interesting to investigate whether our method can be applied to testing precision matrices in the presence of heavy-tailed data, which is often modeled by the elliptical distribution family. We leave this to future work.
% \cite{CQY16} implemented such data-dependent procedure in the statistical inference on the ultra high-dimensional precision matrix with more complex time series data.

\vspace{-10pt}
\section{Supplementary Materials}

Web Appendices, which include proofs of the main theorems and additional numerical results referenced in Sections \ref{method.sec}, \ref{simulation.sec} and \ref{real} are available with this paper on the Biometrics website on Wiley Online Library.\vspace*{-8pt}

%\section*{Acknowledgement}

%=================================================================================================%

\backmatter

%  This section is optional.  Here is where you will want to cite
%  grants, people who helped with the paper, etc.  But keep it short!

\section*{Acknowledgements}

The authors thank the AE and two anonymous referees for constructive comments and suggestions which have improved the presentation of the paper. Jinyuan Chang was supported in part by the Fundamental Research Funds for the Central Universities (Grant No. JBK160159, JBK150501, JBK140507, JBK120509), NSFC (Grant No. 11501462), the Center of Statistical Research at SWUFE and the Australian Research Council. Wen Zhou was supported in part by NSF Grant IIS-1545994. Lan Wang was supported in part by NSF Grant 
NSF DMS-1512267. \vspace*{-8pt}

%  If your paper refers to supplementary web material, then you MUST
%  include this section!!  See Instructions for Authors at the journal
%  website http://www.biometrics.tibs.org

%  Here, we create the bibliographic entries manually, following the
%  journal style.  If you use this method or use natbib, PLEASE PAY
%  CAREFUL ATTENTION TO THE BIBLIOGRAPHIC STYLE IN A RECENT ISSUE OF
%  THE JOURNAL AND FOLLOW IT!  Failure to follow stylistic conventions
%  just lengthens the time spend copyediting your paper and hence its
%  position in the publication queue should it be accepted.

%  We greatly prefer that you incorporate the references for your
%  article into the body of the article as we have done here
%  (you can use natbib or not as you choose) than use BiBTeX,
%  so that your article is self-contained in one file.
%  If you do use BiBTeX, please use the .bst file that comes with
%  the distribution.  In this case, replace the thebibliography
%  environment below by
%
%  \bibliographystyle{biom}
% \bibliography{mybibilo.bib}

\begin{thebibliography}{99}

%\bibitem[Ashburner et al.(2000)]{A00}
%Ashburner, M., Ball, C., Blake, J., Botstein, D., Butler, H., Cherry, J., Davis, A., Dolinski, K., Dwight, S., Eppig, J., Harris, M., Hill, D., Issel-Tarver, L., Kasarskis, A., Lewis, S., Matese, J., Richardson, J., Ringwald, M., Rubin, G., and Sherlock, G. (2000).
%Gene ontology: Tool for the unification of biology.
%{\it  Nature Genetics} {\bf 25}, 25--29.

\bibitem[{Anderson(2003)}]{Anderson_2003}
{Anderson, T. W.} (2003).
\textit{{An Introduction to Multivariate Statistical Analysis}}.
3rd edition. New York: Wiley-Interscience.

\bibitem[{Benjamini and Yekutieli(2001)}]{BY2001}
{Benjamini, Y. and Yekutieli, D.} (2001).
{The control of the false discovery rate in multiple testing under dependency.}
{\it The Annals of Statistics} {\bf 29}, 1165--1188.

\bibitem[Cai et al.(2013)]{CaiLiuXia_2013}
{Cai, T. T., Liu, W., and Xia, Y.} (2013).
{Two-sample covariance matrix testing and support recovery in high-dimensional and sparse settings.}
{\it Journal of the American Statistical Association} {\bf 108}, 265--277.


\bibitem[Caragea and Smith(2007)]{CS07}
{Caragea, P. and Smith, R.} (2007).
{Asymptotic properties of computationally efficient alternative estimators for a class of multivariate normal models}.
{\it Journal of Multivariate Analysis} {\bf 98}, 1417--1440.


%\bibitem[{Chang, Qiu and Yao(2016)}]{ChangQiu_2015}
%Chang, J., Qiu, Y. and Yao, Q. (2016). Statistical inference for large precision matrices with dependent data and their applications. {\sl Preprint}.


\bibitem[{Chang et al.(2014)}]{ChangZhouZhou_2014}
{Chang, J., Zhou, W., and Zhou, W.-X.} (2014).
{Simulation-based hypothesis testing of high dimensional means under covariance heterogeneity.}
Available at {\it arXiv:1406.1939.}


\bibitem[{Chen and Qin(2010)}]{ChenQin_2010}
Chen, S. X. and Qin, Y. (2010).
A two-sample test for high-dimensional data with applications to gene-set testing.
{\it The Annals of Statistics} {\bf 38}, 808--835.


\bibitem[{Chernozhukov et al.(2013)}]{ChernozhukovChetverikovKato_2013}
{Chernozhukov, V., Chetverikov, D., and Kato, K.} (2013).
{Gaussian approximations and multiplier bootstrap for maxima of sums of high-dimensional random vectors.}
{\it The Annals of Statistics} {\bf 41}, 2786--2819.


%\bibitem[Choudhry et al.(2007)]{CSB2007}
%{Choudhry, S., Seibold, M. A., and Borrell, L. N.} (2007).
%{Dissecting complex diseases in complex populations: Asthma in Latino Americans.}
%{\it Proceedings of the American Thoracic Society} {\bf 4}, 226--233.

\bibitem[{de la Fuente(2010)}]{delaFuente2010}
{de la Fuente, A.} (2010).
From differential expression to differential networking -- identification of dysfunctional regulatory networks in diseases.
{\it Trends in Genetics} {\bf 26}, 326--333.

\bibitem[D'haeseleer(2005)]{Dhaeseleer05}
{D'haeseleer, P.} (2005).
How does gene expression clustering work?
{\it Nature Biotechnology} {\bf 23}, 1499--1501.

%\bibitem[{Fortunato et al.(2010)}]{Fortunato10}
%Fortunato, R. S., Lima de Souza, E. C., Ameziane-el Hassani, R., Boufraqech, M., Weyemi, U., Talbot, M., Lagente-Chevallier, O., de Carvalho, D. P., Bidart, J. M., Schlumberger, M., and Dupuy, C. (2010).
%Functional consequences of dual oxidase-thyroperoxidase interaction at the plasma membrane.
%{\it The Journal of Clinical Endocrinology and Metabolism} {\bf 95}, 5403--5411.

\bibitem[Gentleman et al.(2005)]{G2005}
Gentleman, R., Irizarry, R. A., Carey, V. J., Dudoit, S., and Huber, W. (2005).
\textit{Bioinformtics and Computational Biology Solutions Using R and Bioconductor}. New York: Springer-Verlag.


%\bibitem[{Ho et al.(2008)}]{Ho08}
%{Ho, J. W., Stehani, M., dos Remedios, C. G., and Charleston, M. A.} (2008).
%{Differential variability analysis of gene expression and its application to human diseases.}
%{\it Bioinformatics} {\bf 24}, 390--398.

%\bibitem[Hubner et al.(2005)]{H05}
%Hubner, N., Wallace, C. A., Zimdahl, H., Petretto, E., Schulz, H., Maciver, F., Mueller, M., Hummel, O., Monti, J., Zidek, V., Musilova, A., Kren, V., Causton, H., Game, L., Born, G., Schmidt, S., M\"{u}ller, A., Cook, S. A., Kurtz, T. W., Whittaker, J., Pravenec, M., and Aitman, T. J. (2005). Integrated transcriptional profiling and linkage analysis for identification of genes underlying disease. {\it  Nature Genetics} {\bf 37}, 244--253.

\bibitem[Katsani et al.(2014)]{K2014}
Katsani, K. R., Irimia, M., Karapiperis, C., Scouras, Z. G., Blencowe, B. J., Promponas, V. J., and Ouzounis, C. A. (2014).
Functional genomics evidence unearths new moonlighting roles of outer ring coat nucleoporins.
{\it Scientific Reports} {\bf 4}, 4655.


\bibitem[{Li and Chen(2012)}]{LiChen_2012}
{Li, J. and Chen, S. X.} (2012).
Two-sample tests for high-dimensional covariance matrices.
{\it The Annals of Statistics} {\bf 40}, 908--940.

\bibitem[Liu et al.(2008)]{LiuLinShao_2008}
{Liu, W., Lin, Z. Y. and Shao, Q.-M.} (2008).
The asymptotic distribution and Berry-Esseen bound of a new test for independence in high dimension with an application to stochastic optimization.
{\it The Annals of Applied Probability} {\bf 18}, 2337--2366.

\bibitem[Markitsis and Lai(2010)]{ML(2010)}
{Markitsis, A. and Lai, Y.} (2010).
{A censored beta mixture model for the estimation of the proportion of non-differentially expressed genes.}
\textit{Bioinformatics} \textbf{26}, 640--646.

%\bibitem[{Nettleton et al.(2006)}]{Nettleton06}
%{Nettleton, D., Hwang, J. T. G., Caldo, R. A., and Wise, R. P.} (2006).
%Estimating the number of true null hypotheses from a histogram of $p$ values.
%{\it Journal of Agriculture, Biological, and Environmental Statistics} {\bf 11}, 337--356.

\bibitem[{Schott(2007)}]{Schott_2007}
{Schott, J. R.} (2007).
A test for the equality of covariance matrices when the dimension is large relative to the sample size.
{\it Computational Statistics and Data Analysis} {\bf 51}, 6535--6542.

\bibitem[{Sharan et al.(2002)}]{SES2002}
{Sharan, R., Elkon, R., and Shamir, R.} (2012).
Cluster analysis and its applications to gene expression data.
{\it Ernst Schering Research Foundation Workshop} {\bf 38}, 83--108.

\bibitem[{Srivastava and Yanagihara(2010)}]{SrivastavaYanagihara_2010}
{Srivastava, M. S. and Yanagihara, H.} (2010).
Testing the equality of several covariance matrices with fewer observations than the dimension.
{\it Journal of Multivariate Analysis} {\bf 101}, 1319--1329.

\bibitem[van der Vaart and Wellner(1996)]{VW96}
van der Vaart, A. W. and Wellner, J. A. (1996).
{\it Weak Convergence and Empirical Processes: With Applications to Statistics.} New York: Springer.

\bibitem[Voraphani et al.(2014)]{V2014}
{Voraphani, N., Gladwin, M. T., Contreras, A. U., Kaminski, N., Tedrow, J. R., Milosevic, J., Bleecker, E. R., Meyers, D. A., Ray, A., Ray, P., Erzurum, S. C., Busse, W. W., Zhao, J., Trudeau, J. B., and Wenzel, S. E.} (2014).
An airway epithelial iNOS-DUOX2-thyroid peroxidase metabolome drives Th1/Th2 nitrative stress in human severe asthma.
{\it Mucosal Immunology} {\bf 7}, 1175--1185.

\bibitem[{Wolen and Miles(2012)}]{WM2012}
{Wolen, A. R. and Miles, M. F.} (2012).
Identifying gene networks underlying the neurobiology of ethanol and alcoholism.
{\it Alcohol Research: Current Reviews} {\bf 34}, 306--317.

%\bibitem[World Health Organization(2013)]{who}
%{World Health Organization.} (2013).
%\newblock {\it World Health Organization Fact Sheet No 307: Asthma.} \url{http://www.who.int/mediacentre/factsheets/fs307/en/}

\bibitem[{Wagaman and Levina(2009)}]{WL2008}
{Wagaman, A. S. and Levina, E.} (2009).
Discovering sparse covariance structures with the Isomap.
{\it Journal of Computational and Graphical Statistics} {\bf 18}, 551--572.

\bibitem[Wang et al.(2015)]{WPL2015}
Wang, L., Peng, B., and Li., R. (2015).
{A high-dimensional nonparametric multivariate test for mean vector.}
{\it Journal of the American Statistical Association} {\bf 110}, 1658--1669.

\bibitem[Xia et al.(2015)]{XTT2015}
Xia, Y., Cai, T., and Cai, T. T. (2015).
Testing differential networks with applications to the detection of gene-gene interactions.
{\it Biometrika} {\bf 94}, 247--266.

\bibitem[{Yi et al.(2007)}]{YST2001}
{Yi, G., Sze, S.-H., and Thon, M.} (2007).
Identifying clusters of functionally related genes in genomes.
{\sl Bioinformatics} {\bf 23}, 1053--1060.

%\bibitem[Yvert et al.(2003)]{Y03}
%Yvert, G., Brem, R. B., Whittle, J., Akey, J. M., Foss, E., Smith, E. N., Mackelprang, R., and Kruglyak, L. (2003).
%Trans-acting regulatory variation in Saccharomyces cerevisiae and the role of transcription factors.
%{\it  Nature Genetics} {\bf 35}, 57--64.
\end{thebibliography}

%=================================================================================================%

%\appendix

%  To get the journal style of heading for an appendix, mimic the following.

%\section{}
%\subsection{Title of appendix}

%Put your short appendix here.  Remember, longer appendices are
%possible when presented as Supplementary Web Material.  Please
%review and follow the journal policy for this material, available
%under Instructions for Authors at \texttt{http://www.biometrics.tibs.org}.

\label{lastpage}

\end{document}